\documentclass[aps,twocolumn,10pt,prc,floatfix,showpacs,preprintnumbers,amsmath,amssymb,nofootinbib,superscriptaddress]{revtex4-1}

\usepackage[normalem]{ulem}
\usepackage{wasysym}
\usepackage{color}
\usepackage{graphicx}
\usepackage{dcolumn}    
\usepackage{multirow, booktabs }
\usepackage{comment}
\usepackage{soul}

\usepackage{physics, mathtools, braket}
\usepackage [ english ]{ babel }

\usepackage{amsmath,amsfonts,amssymb,amsthm}  
\usepackage{latexsym}
\usepackage{bbold}

\usepackage{bigstrut}
\usepackage{CJK}
\usepackage[pdfstartview=FitH,
            CJKbookmarks=true,
            bookmarksnumbered=true,
            bookmarksopen=true,
            colorlinks,
            pdfborder=001,
            linkcolor=blue,
            anchorcolor=blue,
            citecolor=blue,
            urlcolor=blue,
            ]{hyperref}

\begin{document}

\newcommand{\IdentityMat}{\mathbb{1}}
\newcommand{\Sch}{ Schr\"{o}dinger }
\newcommand{\etal}{\textit{et al.} }

\newcommand{\FM}[1]{{\color{magenta} #1}}

\newcommand\rnumber{\operatorname{r-number}}

\title{Structure and dynamics of open-shell nuclei from spherical coupled-cluster theory  }

\thanks{This manuscript has been authored in part by UT-Battelle, LLC, under contract DE-AC05-00OR22725 with the US Department of Energy (DOE). The US government retains and the publisher, by accepting the article for publication, acknowledges that the US government retains a nonexclusive, paid-up, irrevocable, worldwide license to publish or reproduce the published form of this manuscript, or allow others to do so, for US government purposes. DOE will provide public access to these results of federally sponsored research in accordance with the DOE Public Access Plan (http://energy.gov/downloads/doe-public-access-plan).}

\author{Francesco Marino}
\email{frmarino@uni-mainz.de}
\affiliation{Institut f\"{u}r Kernphysik and PRISMA+ Cluster of Excellence, Johannes Gutenberg-Universit\"{a}t Mainz, 55128 Mainz, Germany}

\author{Francesca Bonaiti}
\email{bonaiti@frib.msu.edu}
\affiliation{ Facility for Rare Isotope Beams, Michigan State University, East Lansing, MI 48824, USA}
\affiliation{Physics Division, Oak Ridge National Laboratory, Oak Ridge, TN 37831, USA}

\author{Sonia Bacca}
\email{s.bacca@uni-mainz.de}
\affiliation{Institut f\"{u}r Kernphysik and PRISMA+ Cluster of Excellence, Johannes Gutenberg-Universit\"{a}t Mainz, 55128 Mainz, Germany}
\affiliation{Helmholtz-Institut Mainz, Johannes Gutenberg-Universität Mainz, D-55099 Mainz, Germany}

\author{Gaute Hagen}
\email{hageng@ornl.gov}
\affiliation{Physics Division, Oak Ridge National Laboratory, Oak Ridge, TN 37831, USA}
\affiliation{Department of Physics and Astronomy, University of Tennessee, Knoxville, TN 37996, USA}

\author{Gustav R. Jansen}
\email{jansengr@ornl.gov}
\affiliation{National Center for Computational Sciences, Oak Ridge National Laboratory, Oak Ridge, TN 37831, USA}
\affiliation{Physics Division, Oak Ridge National Laboratory, Oak Ridge, TN 37831, USA}

\begin{abstract}
We extend the spherical coupled-cluster \textit{ab initio} method for open-shell nuclei where two nucleons are removed from a shell subclosure.
Following the recent implementation of the two-particle-attached approach [Phys.~Rev.C 110 (2024) 4, 044306], we focus on the two-particle-removed method.  
Using the equations-of-motion framework, we address both nuclear structure and dipole response functions by coupling coupled-cluster theory with the Lorentz integral transform technique.
We perform calculations using chiral interactions, including three-nucleon forces, and estimate many-body uncertainties by comparing different coupled-cluster truncation schemes.
We validate our approach by studying ground-state energies, excited states, and electric dipole polarizabilities in the oxygen and calcium isotopic chains.
For binding energies and selected low-lying excited states, we achieve an accuracy comparable to that of the established closed-shell coupled-cluster theory and generally agree with experiment.
Finally, we underestimate experimental data for electric dipole polarizabilities, particularly in calcium isotopes.
\end{abstract}

\maketitle

\section{Introduction}
\label{sec: intro}

The goal of \textit{ab initio} nuclear theory~\cite{Hergert2020,papenbrock2024,Ekstrom2023,computational_nuclear,BaccaPastore2014} is to understand the complex emergent phenomenology of atomic nuclei based on a fundamental description of nuclear interactions at the scale of nucleons and pions, such as that provided by chiral effective field theory (EFT)~\cite{Epelbaum2006Review,Machleidt2011,Epelbaum2024,MACHLEIDT2024104117}.
The development of advanced many-body methods is crucial for improving the accuracy of predictions, quantifying theoretical uncertainties, expanding the reach of calculations beyond light and closed-shell nuclei, and studying new observables.

Various techniques exist to solve the many-nucleon \Sch equation~\cite{Hergert2020,computational_nuclear}.~The no-core shell model (NCSM)~\cite{Barrett2013NoCore}, hyperspherical harmonics~\cite{Marcucci:2019hml,PhysRevLett.130.152502,LiMuli:2024puj}, Green's function Monte Carlo~\cite{Lovato:2017cux,King:2022zkz}, and lattice EFT~\cite{LEE2009117,lee2025latticeeffectivefieldtheory} are nearly-exact techniques for  light nuclei. 
For heavier systems, systematically improvable many-body expansion methods, such as self-consistent Green's functions (SCGF) theory~\cite{Barbieri2004,Soma2020}, the in-medium similarity renormalization group~\cite{Hergert2016Imsrg,Stroberg2021}, and the coupled-cluster (CC) method~\cite{Bartlett2007,ShavittBartlett,Hagen2014Review,Hagen_2016_rev}, are employed, thanks to their favorable computational cost, which scales polynomially with the number of nucleons.

Thanks to its flexibility and excellent accuracy, the CC method has played a pivotal role in \textit{ab initio} nuclear theory, allowing, e.g., to address properties of medium-mass and heavy-mass nuclei~\cite{PbAbInitio}.
Over time, it has evolved into a powerful tool for predicting ground state (g.s.)~properties (binding energies, charge densities, and weak form factors~\cite{HagenNature2015,Payne2019}), rotational bands and electromagnetic transitions in deformed nuclei using symmetry breaking and restoration techniques ~\cite{Novario2020,Hagen2022,Sun2025}, and, through the equations-of-motion (EOM) framework~\cite{Bartlett2007,Krylov2008}, for computing low-lying spectra and electroweak responses (see e.g.~\cite{Bacca2013,Bacca2014,Sobczyk2021,Sobczyk:2023mey,Sobczyk:2023sxh}).~Pioneering calculations~\cite{Bacca2013,Bacca2014} coupling the EOM formalism with the Lorentz integral transform (LIT) technique~\cite{Efros1994,Efros_2007,Bacca2014} have provided the first \textit{ab initio} calculations of nuclear response functions beyond light nuclei, in regions where microscopic computations have been traditionally performed with more phenomenological energy density functional methods~\cite{Paar_2007,rocamaza2018,colo2020}.
More recently, electromagnetic responses have also been examined using NCSM~\cite{Stumpf2017} (in nuclei with $A\le 24$), SCGF~\cite{Raimondi2019}, the projected generator coordinate method (PGCM)~\cite{Porro:2024pdn,Porro:2024vlc}, and quasi-particle finite-amplitude method~\cite{Beaujeault2023}.

While applications to closed-shell nuclei are relatively well consolidated, attacking open-shell systems is an ongoing challenge for \textit{ab initio} theory.
Gorkov SCGF~\cite{Soma2020,Soma2011,Soma2020Chiral}, for instance, employs a superfluid formalism to describe even-even nuclei, while PGCM~\cite{Frosini2021mrIII} and IMSRG extensions~\cite{Yao2020,Belley2024} can describe spectra of deformed nuclei using multi-reference approaches.
PGCM~\cite{Porro:2024pdn} and NCSM~\cite{Stumpf2017} have also been applied to response functions.
Symmetry-breaking approaches followed by symmetry restoration techniques are currently at the center of investigation in CC theory, as demonstrated by recent calculations performed with Bogoliubov CC (BCC)~\cite{Signoracci2015,Tichai2024} and deformed CC~\cite{Novario2020,Hagen2022,Hu2024,Sun2025,Sun2025Odd}.
In the former, pairing correlations are included self-consistently in the reference state by starting from a Hartree-Fock-Bogoliubov solution, allowing the study of open-shell even-even nuclei. 
This approach breaks particle-number conservation in the reference state.
In the latter, single-reference CC calculations are performed on top of a deformed HF state, which breaks rotational invariance, allowing, in principle, the study of generic deformed nuclei. 
These approaches are very general and promising. 
While less critical for size-extensive quantities such as binding energies~\cite{papenbrock2024}, the broken symmetries should eventually be restored when computing excited states and their transitions~\cite{Hagen2022,Duguet_2015,Duguet_2017,Sun2025}.
However, the need to perform particle-number and/or angular-momentum projection~\cite{Hagen2022,Sheikh_2021} significantly affects both the complexity and the computational cost of BCC and deformed CC, respectively.

A symmetry-preserving alternative for open-shell nuclei is provided by the EOM method~\cite{Krylov2008,Bartlett2007,Gour2006,Jansen2011}.
In this approach, an open-shell nucleus near a sub-shell closure is treated as an excited state of its closed-shell neighbor.
The calculation proceeds in two steps: first, a standard CC computation determines the g.s.~of the reference core; second, an EOM equation is solved for the excitation amplitudes describing the open-shell nucleus.
The so-called particle-attached/particle-removed EOM schemes are conceptually simple and computationally efficient, and are well-suited for states with a relatively simple structure~\cite{Jansen2011,Krylov2008}.
This has been demonstrated in the one-particle-attached/removed EOM for one-nucleon valence nuclei~\cite{Gour2006,Hagen2014Review} and the two-particle-attached (2PA) method~\cite{Jansen2011,Jansen2013}. 
Importantly, a 2PA-LIT-CC technique has been introduced recently~\cite{Bonaiti2024}, enabling calculations of nuclear response functions in open-shell nuclei. 
This method has been applied to the electric dipole response function, and specifically to compute the electric dipole polarizability $\alpha_D$~\cite{Bonaiti2024, Brandherm2024}. 

In this paper, we extend the approach by developing the two-particle-removed (2PR) CC technique~\cite{Jansen2011,Piecuch2013,Marino2024Nsd}, which enables the study of nuclei that have two nucleons removed from a shell closure.
We present the formalism for bound states (2PR-EOM-CC) and response functions (2PR-LIT-CC) in detail (Sec.~\ref{sec: Formalism}).
We validate the method through extensive numerical calculations of ground-state energies, selected excited states, and electric dipole polarizabilities (Sec.~\ref{sec: validating 2pr}).
By combining the different EOM schemes, we provide comprehensive predictions for binding energies and electric dipole polarizabilities in the oxygen and calcium isotope chains.
Finally, we discuss conclusions and future developments in Sec.~\ref{sec: conclusions}.

\section{Formalism}
\label{sec: Formalism}

The intrinsic nuclear Hamiltonian~\cite{Hagen2014Review}
\begin{align}
& \hat{H}=\left(1-\frac{1}{A^*}\right) \sum_{i=1}^A \frac{ \hat{p}_i^2}{2 m}  \nonumber \\
& + \sum_{i<j=1}^A \left( \hat{v}_{i j} -\frac{\mathbf{\hat{p}}_i \cdot \mathbf{\hat{p}}_j}{m A^*}\right) +
\sum_{i<j<k} \hat{w}_{ijk}
\end{align}
serves as the starting point of our calculations.
Here, $\hat{v}_{ij}$ and $\hat{w}_{ijk}$ are the two- (NN) and three-nucleon (3N) interactions, respectively, $\hat{\mathbf{p}}_i$ is the momentum of the $i$-th nucleon, $m$ is the nucleon mass, $A$ is the number of nucleons in the reference state, and $A^{*}$ is the mass of the target nucleus. 
For closed-shell CC, $A^{*} = A$, while in the 2PR case, $A^{*} = A-2$.
Applying this mass shift ensures that the correct kinetic energy is used for the target nucleus.

\subsection{Single-reference coupled-cluster}
\label{sec: gs coupled cluster}
The single-reference coupled-cluster framework~\cite{Hagen2014Review,ShavittBartlett} employs an exponential parametrization of the many-body ground state,
\begin{align}
    \label{eq: cc gs ansatz}
    \ket{\Psi_0} = e^{ \hat{T} } \ket{ \Phi_0 },
\end{align}
where $\ket{ \Phi_0 }$ is a reference state, commonly obtained from a variational mean-field solution, e.g., a Hartree-Fock (HF) Slater determinant.
Correlations are built on top of $\ket{ \Phi_0 }$ by the action of $e^{ \hat{T} }$, where the cluster operator $\hat{T}$ is expanded as a combination of $n$-particle-$n$-hole ($n$p-$n$h) excitation operators $T_n$, i.e.,
\begin{align}
    \hat{T} = \sum_{n=1}^{A} \hat{T}_n =  \hat{T}_1 + \hat{T}_2 + ... + \hat{T}_A,
\end{align}
and the Taylor expansion is used to represent the exponential function.

In practice, to limit the complexity and the computational cost of the method, the cluster operator has to be truncated up to a certain number of $ph$ excitations. 
A simple truncation, CC at the singles and doubles level (CCSD), includes terms up to $2p$-$2h$ excitations, i.e., $T = T_1 + T_2$.
The $T$ operators read~\cite{Hagen2014Review}
\begin{subequations}
\begin{align}
    & T_1 = \sum_{ai} t^{a}_{i} c_a^{\dagger} c_i\, , \\
    & T_2 = \frac{1}{4} \sum_{abij} t^{ab}_{ij} c_a^{\dagger} c_b^{\dagger} c_j c_i\,  ,
\end{align}
\end{subequations}
where indices $i,j$ and $a,b$ denote single-particle (s.p.) states that are occupied (holes) and unoccupied (particles) in the reference state, respectively.
The amplitudes are antisymmetric w.r.t.~the permutation of indices within the ket or bra side. 

We work within the normal-ordered two-body approximation~\cite{Hagen2007,roth2012,Hebeler2021}. 
To this end, we introduce the Hamiltonian normal-ordered wrt.~$\ket{ \Phi_0 }$ by $\hat{H}_{N} = \hat{H} - E_{0}^{\rm ref}$, with the reference energy given by $E_{0}^{\rm ref} = \mel{\Phi_0}{\hat{H}}{\Phi_0}$.
In this scheme, 3N forces contribute through the one- and two-body matrix elements of $\hat{H}_{N}$, while explicit 3N operators are neglected. 
Then, we consider the~\Sch equation $\hat{H} \ket{\Psi_0} = E_{0} \ket{\Psi_0} $, where $E_{0}$ is the CC g.s.~energy.
Inserting Eq.~\eqref{eq: cc gs ansatz}, the key equation for the coupled-cluster g.s. is found:
\begin{equation}
    \label{eq: CC basic eq}
    \Bar{H} \ket{ \Phi_0 } = \Delta E_0 \ket{ \Phi_0 },
\end{equation}
where the similarity-transformed Hamiltonian
\begin{align}
    \Bar{H} = e^{-\hat{T}} \hat{H}_N e^{\hat{T}} = ( \hat{H}_N e^{\hat{T}} )_{C}
\end{align}
has been introduced, and $\Delta E_0 = E_{0} - E_{0}^{\rm ref}$.
The subscript $C$ indicates that only connected diagrams must be considered. 
By projecting Eq.~\eqref{eq: cc gs ansatz} onto the reference state, the CC energy can be evaluated as 
\begin{align}
    \label{eq: cc energy}
    \Delta E_0 = \mel{ \Phi_0 }{ \Bar{H} }{ \Phi_0 },
\end{align}
while projecting onto the excited Slater determinants $\ket{ \Phi_{i}^{a} } = c_{a}^{\dagger} c_i \ket{\Phi_0} $, $\ket{ \Phi_{ij}^{ab} } = c_{a}^{\dagger} c_{b}^{\dagger} c_{j} c_{i} \ket{\Phi_0} $ etc., a set of (non-linear) equations for the amplitudes is determined.
In the CCSD approximation,
the amplitude equations are given by
\begin{align}
    \mel{ \Phi_{i}^{a} }{ \Bar{H} }{ \Phi_0 } & = 0, \\
    \mel{ \Phi_{ij}^{ab} }{ \Bar{H} }{ \Phi_0 } &= 0.
\end{align}
Additional $3p$-$3h$ (triples) correlations can be included in an approximate fashion to achieve greater precision using e.g. the CCSDT-1 or CCSDT-3 approaches~\cite{lee1984,noga1987}. 

A peculiarity of the CC method is that the similarity-transformed Hamiltonian is not Hermitian~\cite{ShavittBartlett}. As a consequence, to compute expectation values or response functions~\cite{Bacca2013,Miorellithesis}, right and left g.s.~eigenvectors are both needed.
In particular, the left eigenstate $\bra{\Psi_0} $ can be written down as
\begin{align}
    \label{eq: left gs}
    \bra{\Psi_0} = \bra{\Phi_0} (1 + \hat{ \Lambda })e^{- \hat{T }},
\end{align} where $\hat{ \Lambda } = \hat{ \Lambda }_1 + \hat{ \Lambda }_2 + ...$ is a de-excitation operator, whose amplitudes constitute an additional unknown of the problem~\cite{ShavittBartlett}.

\subsection{Equation-of-motion coupled-cluster}
\label{sec: eom CC}

The single-reference CC framework has been applied successfully to the g.s.~of closed-(sub)shell nuclei, see e.g.~\cite{Hagen2014Review,HagenNature2015,PbAbInitio,Hagen2016}. Excited states can be efficiently computed within the equations-of-motion (EOM) coupled-cluster approach~\cite{Stanton1993,Bartlett2007,Krylov2008,Hagen2014Review,Hagen2010}.
At the same time, the EOM approach provides a solution to the problem of studying open-shell nuclei.
The EOM technique does not break any symmetries, and provides an efficient and conceptually simple way of studying the spectrum and the response of open-shell systems in the vicinity of a closed-shell nucleus~\cite{Jansen2011,Musial2003,Gour2006,Krylov2008}.
Given a closed-shell nucleus with g.s.~wave function $\ket{\Psi_0}$, within the EOM its neighboring open-shell nuclei with mass $A^{*} = A \pm k$ are interpreted as excited states of the closed-shell nucleus itself.

EOM-CC is a two-stage method. First, the g.s.~($\hat{T}$ and $\hat{\Lambda}$ amplitudes) of a given spherical nucleus is determined as in Sec.~\ref{sec: gs coupled cluster}.
Then, one postulates that target eigenstates $\ket{ \Psi_f }$ of the Hamiltonian can be approximated by acting linearly on the g.s.~with an excitation operator.
Within this configuration-interaction ansatz, the right and left eigenstates are given by
\begin{align}
    \label{eq: eom right ansatz}
    \ket{ \Psi_f } = \hat{ R }_{f}  \ket{\Psi_0} = \hat{ R }_{f} e^{ \hat{T} } \ket{ \Phi_0 },
    \\ 
    \label{eq: eom left ansatz}
    \bra{ \Psi_f } = \bra{ \Psi_0 } \hat{L}_{f} = \bra{ \Phi_0 }  \hat{L}_{f}e^{ - \hat{T} } .
\end{align}
By construction, the amplitudes $\hat{ R }_{f}$ and $\hat{ L }_{f}$ are normalized such that $
\mel{\Phi_0}{ \hat{L}_f \hat{R}_{ f^\prime } }{ \Phi_0 } = \delta_{f f^\prime}$.
Inserting the wave functions from Eqs.~\eqref{eq: eom right ansatz} and \eqref{eq: eom left ansatz} into the~\Sch equation, the following eigenvalue equations are found,
\begin{align}
    \label{eq: right eom eq}
    ( \Bar{H} \hat{ R }_{f} )_{C} \ket{ \Phi_0 } & =
    \omega_{f} \hat{ R }_{f} \ket{ \Phi_0 }, \\
    \label{eq: left eom eq}
    \bra{ \Phi_0 } \hat{L}_{f} \Bar{H} &= \bra{ \Phi_0 } \hat{L}_{f} \omega_f,
\end{align}
where $\omega_f = \Delta E_f - \Delta E_{0}$ denotes the excitation energy of the states $f$ of the target nucleus wrt.~the ground state. 
Both cases involve evaluating the product between the transformed Hamiltonian and the excitation operator. However, the right eigenvector $\hat{ R }_{f}$ is connected to $\Bar{H}$, while the left eigenvector $\hat{ L }_{f}$ may not be~\cite{Bartlett2007}. 

In EOM-CC, the excited states of the $A$-particle system are parametrized in terms of $np-nh$ configurations. For example, at the singles and doubles level, the right and left operators read
\begin{align}
    \hat{ R }_{f} &= r_0 + \sum_{ai} r^{a}_{i} c^{\dagger}_{a} c_i + \frac{1}{4} \sum_{abij} r^{ab}_{ij} c^{\dagger}_{a} c^{\dagger}_{b} c_{j} c_{i}, 
    \\
    \hat{ L }_{f} &= l_0 + \sum_{ai} l^{i}_{a} c^{\dagger}_{i} c_{a} + \frac{1}{4} \sum_{abij} l_{ab}^{ij} c^{\dagger}_{i} c^{\dagger}_{j} c_{b} c_{a}, 
\end{align}
respectively.
If the g.s.~is targeted, in particular, $\hat{R}_0 = 1$ and $\hat{L}_{0} = 1 + \hat{\Lambda}$.
The amplitudes of $\hat{R}_f$ and $\hat{L}_f$ are determined by casting Eqs.~\eqref{eq: right eom eq} and \eqref{eq: left eom eq} in matrix form and using the Arnoldi algorithm to solve for the first few low-lying eigenvalues~\cite{Hagen2014Review}.

We now focus on the two-particle-removed (2PR) formalism~\cite{Jansen2011,Piecuch2013}, which applies to nuclei that differ by two holes from a shell closure.
We assume that an ansatz similar to that of Eqs.~\eqref{eq: eom right ansatz} and \eqref{eq: eom left ansatz} applies to the states of the 2PR target nucleus.
Thus, we first compute the g.s.~of a $A$-nucleon closed-shell reference with the mass shift $A^{*}=A-2$ (see Ref.~\cite{Jansen2011}), yielding a CC correlation energy $\Delta E_0^{*}$.
At variance with the conventional EOM, the excitation operator, which connects the closed-shell core to the states of the open-shell nucleus, does not conserve the particle number, but rather removes two fermions. 
At leading order, it includes $0p$-$2h$ configurations, which must be complemented by at least $1p$-$3h$ contributions.
A treatment of 2PR-EOM at the $2p$-$4h$ level, discussed in quantum chemistry in Refs.~\cite{Shen19032014,Piecuch2025}, is left to future work.
Then, the 2PR operators read~\cite{Jansen2011,Piecuch2013}
\begin{align}
    \label{eq: eom 2pr right}
    & \hat{R}_{f}^{(A-2)} = 
    \frac{1}{2} \sum_{ij} r_{ij} c_j c_i + \frac{1}{6} \sum_{ijka} r_{ijk}^{a} c_a^\dagger c_k c_j c_i, \\
    \label{eq: eom 2pr left}
    & \hat{L}_{f}^{(A-2)} = 
    \frac{1}{2} \sum_{ij} l^{ij} c^{\dagger}_{i} c^{\dagger}_{j} + \frac{1}{6} \sum_{ijka} l^{ijk}_{a} c^{\dagger}_{i} c^{\dagger}_{j} c^{\dagger}_{k} c_{a} .
\end{align}
The 2PR-EOM equations are given by
\begin{align}
    \label{eq: right 2pr eom}
    ( \Bar{H} \hat{ R }^{(A-2)}_{f} )_{C} \ket{ \Phi_0 } &=
    \omega_{f}^{(A-2)} \hat{ R }^{(A-2)}_{f} \ket{ \Phi_0 }, \\
    \label{eq: left 2pr eom}
    \bra{ \Phi_0 } \hat{L}_{f}^{(A-2)} \Bar{H} &= \bra{ \Phi_0 } \hat{L}_{f}^{(A-2)} \omega_f^{(A-2)},
\end{align}
where
$\omega_f^{(A-2)} = \Delta E_f^{(A-2)} - \Delta E_0^{*}$ is the excitation energies of the open-shell nucleus wrt.~the mass-shifted closed-shell reference. 
The diagrams representing the product $\Bar{H} \hat{ R }^{(A-2)}_{f}$ are reported in Appendix~\ref{app: spherical 2pr} and Tab.~\ref{tab: 2pr eom diagrams}.

\subsection{Lorentz integral transform}
\label{sec: lit}
The description of the response of a nucleus to an external probe $\hat{\Theta}$ is encoded in the response function
\begin{align}
    \label{eq: response basic def}
    R(\omega) = \sum_{f} \mel{\Psi_{0}}{ \hat{\Theta}^{\dagger} }{ \Psi_{f}}
    \mel{\Psi_{f}}{ \hat{\Theta} }{ \Psi_{0}}
    \delta( E_{f} - E_{0} - \omega),
\end{align}
where the perturbation connects the ground state $\ket{ \Psi_{0} }$ to the excited states $\ket{ \Psi_{f} }$.
The sum over final states includes both the bound spectrum and a continuum of excited states.
Accessing the continuum spectrum is challenging since it requires evaluating contributions from all possible breakup channels.
The Lorentz integral transform (LIT) technique~\cite{Efros1994,Efros_2007} simplifies the problem by dealing with the continuum in an implicit way.
An integral transform $L(\sigma,\Gamma)$ of the original response function $R(\omega)$ is introduced by means of a convolution with a Lorentzian kernel, namely
\begin{align}
    \label{eq: lit def}
    L(\sigma,\Gamma) &= \int d\omega K_\Gamma(\sigma,\omega) R(\omega) \\
    &= \frac{\Gamma}{\pi} \int d\omega \, \frac{ R(\omega) }{ (\sigma-\omega)^2 + \Gamma^2 } \,,\nonumber
\end{align}
where $K_\Gamma(\sigma,\omega)$ is a Lorentzian function of width $\Gamma$.
$K_\Gamma(\sigma,\omega)$ is a convenient representation of the delta function.
Since it satisfies $\lim_{\Gamma \to 0} K_\Gamma(\sigma,\omega) = \delta(\sigma-\omega)$, the LIT in the limit of vanishing width corresponds to a discretized representation of $R(\omega)$. We dub this quantity as the discretized response function.

Inserting the definition~\eqref{eq: response basic def} into Eq.~\eqref{eq: lit def}, with some manipulation we find that 
\begin{align}
    L(\sigma,\Gamma) &= \frac{\Gamma}{\pi}
    \bra{ \Psi_{0} }
    \hat{\Theta}^{\dagger}
    \frac{1}{ \hat{H} - E_0 - \sigma + i\Gamma}
    \\
    & \times
    \frac{1}{ \hat{H} - E_0 - \sigma - i\Gamma}
    \hat{\Theta}
    \ket{ \Psi_{0} } .
    \nonumber
\end{align}
Defining the variable $z = \sigma + E_0 + i\Gamma$, the above expression can be rewritten as the norm of an auxiliary state $\ket{ \Tilde{\Psi}_0(z) }$,
\begin{align}
    \label{eq: lit as norm}
    L(z) = \frac{\Gamma}{\pi} \innerproduct*{ \Tilde{\Psi}_0(z^{*}) }{ \Tilde{\Psi}_0(z) }\,,
\end{align}
where
\begin{align}
    \ket{ \Tilde{\Psi}_0(z) } = \frac{1}{ \hat{H} - z}
    \hat{\Theta}
    \ket{ \Psi_{0} }.
\end{align}
Since the lhs of Eq.~\eqref{eq: lit as norm} is finite for any $\Gamma$, $\ket{ \Tilde{\Psi}_0(z) }$ satisfies bound-state-like boundary conditions.
Thus, the advantage of the method lies in the fact that computing $L(\sigma,\Gamma)$ reduces to a bound-state-like problem.
The response function can then be recovered from the LIT by a numerical inversion~\cite{Efros1994,Efros_2007}.

The moments of the response functions can also be obtained directly from the LIT in the limit of a small width without the need of any inversion~\cite{Miorelli2016} as
\begin{align}
    \label{eq: moments def}
    m_n = \int d\omega \, \omega^{n} R(\omega) 
    = \lim_{\Gamma \to 0} \int d\sigma \, \sigma^{n} L(\sigma,\Gamma).
\end{align}

\subsection{Electromagnetic response for two-particle-removed nuclei}
\label{sec: lit 2pr cc}
The LIT formalism can be coupled with the 2PR-EOM coupled-cluster ansatz. The resulting 2PR-LIT-CC allows  to address electromagnetic response functions of two-particle-removed open-shell nuclei,
see also Refs.~\cite{Bacca2014,Bonaiti2024,Miorellithesis}.
We evaluate the response function of Eq.~\eqref{eq: response basic def} for an $(A-2)$-body nucleus by assuming the EOM ansatz for the eigenstates of the Hamiltonian.
We use Eqs.~\eqref{eq: eom 2pr left} and \eqref{eq: eom 2pr right} for the left and right wave functions, respectively.
We find
\begin{align}
    \label{eq: response 2pr long}
    & R(\omega) = \sum_{f}
    \mel{\Phi_{0}}{ \hat{L}^{(A-2)}_{0} \overline{ \Theta}^\dagger\hat{R}^{(A-2)}_{f} }{ \Phi_{0} }
    \\ 
    & \times
    \mel{\Phi_{0}}{ \hat{L}^{(A-2)}_{f} \Bar{\Theta} \hat{R}^{(A-2)}_{0} }{ \Phi_{0} }
    \delta( \Delta E_{f}^{(A-2)} - \Delta E_{0}^{(A-2)} - \omega),
    \nonumber
\end{align}
where the similarity-transformed excitation operators have been introduced as
\begin{align}
    & \Bar{\Theta} = e^{ -\hat{T} } \hat{\Theta} e^{ \hat{T} }, \\
    & \overline{ \Theta}^\dagger = e^{ -\hat{T} } \hat{\Theta}^{\dagger} 
    e^{ \hat{T} }.
\end{align}
From now on, the superscript $(A-2)$ will be dropped to simplify the notation.
By adding and subtracting the g.s.~energy of the closed-shell nucleus $\Delta E_{0}^{*}$, the delta function can be written as $\delta( \omega_{f} - \omega_{0} - \omega)$, where $\omega_0 = \Delta E_0 - \Delta E_{0}^{*}$ represents the difference between the g.s.~energies of the open-shell nucleus and its closed-shell neighbour. 
For the response of a closed (sub-)shell nucleus, we trivially have $\omega_0 = 0$.
Then, the response function becomes
\begin{align}
    \label{eq: response 2pr compact}
    & R(\omega) = \sum_{f}
    \mel{\Phi_{0}}{ \hat{L}_{0} \overline{ \Theta}^\dagger\hat{R}_{f} }{ \Phi_{0} }
    \\ 
    & \times
    \mel{\Phi_{0}}{ \hat{L}_{f} \Bar{\Theta} \hat{R}_{0} }{ \Phi_{0} }
    \delta( \omega_{f} - \omega_{0} - \omega )\,.
    \nonumber
\end{align}
Inserting Eq.~\eqref{eq: left eom eq} and using the completeness relation, one can write the response function in operator form as
\begin{align}
    \label{eq: response 2pr operator}
    R(\omega) =
    \mel{ \Phi_{0} }{
    \hat{L}_{0}
    \overline{ \Theta}^\dagger
    \delta( \bar{H} - \omega_0 - \omega)
    \Bar{\Theta}
    \hat{R}_{0}
    }{ \Phi_{0} }\,.
\end{align}
The LIT can now be evaluated from Eq.~\eqref{eq: response 2pr operator}, see also Refs.~\cite{Bonaiti2024,Bacca2014}, as
\begin{align}
    L(\sigma,\Gamma) = \frac{\Gamma}{\pi}
    \mel{ \Phi_0 }{
    \hat{L}_{0}
    \overline{ \Theta}^\dagger
    (\bar{H} - z^{*})^{-1}
    (\bar{H} - z)^{-1}
    \Bar{\Theta}
    \hat{R}_{0}
    }{ \Phi_0 },
\end{align}
where $z = \sigma + \omega_0 + i\Gamma$.
This relation can be rewritten as 
\begin{align}
    \label{eq: lit norm RL}
    L(z) =
    \frac{\Gamma}{\pi} \innerproduct{ \Psi_L(z^{*}) }{ \Psi_R(z) }
\end{align}
in terms of the auxiliary states $\ket{ \Psi_R(z) }$ and $\bra{ \Psi_L(z^{*}) }$, which are defined as the solutions to the following Sch\"odinger-like equations with source terms,
\begin{align}
    \label{eq: aux state 2pr lit right}
    (\bar{H} - z) \ket{ \Psi_R(z) } = \Bar{\Theta}
    \hat{R}_{0} \ket{\Phi_0}, \\
    \label{eq: aux state 2pr lit left}
    \bra{ \Psi_L(z^{*}) } (\bar{H} - z^{*}) = 
    \bra{ \Phi_0 } \hat{L}_{0} 
    \overline{ \Theta}^\dagger.
\end{align}
For later convenience, we introduce the left and right pivots as
\begin{align}
    \label{eq: pivot right}
    \mathbf{S}_R &= \Bar{\Theta}
    \hat{R}_{0} \ket{\Phi_0}, \\
    \label{eq: pivot left}
    \mathbf{S}_L &= \bra{ \Phi_0 } \hat{L}_{0} 
    \overline{ \Theta}^\dagger.
\end{align}
It is also easy to show, e.g., by integrating Eq.~\eqref{eq: response 2pr long}, that the $m_0$ sum rule is given by 
\begin{align}
    m_0 = \mathbf{S}_{L} \cdot \mathbf{S}_{R}.
\end{align}
Next, an ansatz analogous to the 2PR-EOM one used in Eq.~\eqref{eq: eom 2pr right} is postulated for $\ket{ \Psi_R(z) }$, namely we assume
\begin{align}
    \label{eq: 2pr ansatz lit right}
    \ket{ \Psi_R(z) } & = \mathcal{R}(z) \ket{\Phi_0},
    \\ 
    \nonumber \mathcal{R}(z) & =
    \frac{1}{2} \sum_{ij} r_{ij}(z) c_j c_i + \frac{1}{6} \sum_{ijka} r_{ijk}^{a}(z) c_a^\dagger c_k c_j c_i,
\end{align}
and similarly
\begin{align}
    \label{eq: 2pr ansatz lit left}
    \bra{ \Psi_L(z^{*}) } &= \bra{\Phi_0} \mathcal{L}(z^{*}) 
    \\ \mathcal{L}(z^{*}) & =
    \frac{1}{2} \sum_{ij} l^{ij}(z^{*}) c^{\dagger}_{i} c^{\dagger}_{j} + \frac{1}{6} \sum_{ijka} l^{ijk}_{a}(z^{*}) c^{\dagger}_{i} c^{\dagger}_{j} c^{\dagger}_{k} c_{a}.
\end{align}
Then, by plugging Eqs.~\eqref{eq: 2pr ansatz lit right} and \eqref{eq: 2pr ansatz lit left} into Eqs.~\eqref{eq: aux state 2pr lit right} and \eqref{eq: aux state 2pr lit left} for the auxiliary states, one finds
\begin{align}
    \label{eq: right ampl with source}
    (\bar{H} - z) \mathcal{R}(z) \ket{\Phi_0} =
    \Bar{\Theta}
    \hat{R}_{0} \ket{\Phi_0}, \\
    \label{eq: left ampl with source}
    \bra{\Phi_0} \mathcal{L}(z^{*}) (\bar{H} - z^{*}) =
    \bra{ \Phi_0 } \hat{L}_{0} 
    \overline{ \Theta}^\dagger.
\end{align}
These equations for the unknown amplitudes $\mathcal{R}(z)$, $\mathcal{L}(z^{*})$ resemble the EOM equations, from which they differ by the presence of a source term on the r.h.s.~and by a dependence on the $z$ variable.
Fortunately, it is not necessary to solve Eqs.~\eqref{eq: right ampl with source} and \eqref{eq: left ampl with source} for each value of $z$.
In fact, the computational load can be reduced significantly by applying the Lanczos method, as detailed in Refs.~\cite{Miorelli2016,Bonaiti2024,Bacca2014}.

An efficient numerical implementation of the EOM and LIT-CC methods has been achieved by exploiting the NTCL library~\cite{NTCL}, which provides a high-level interface for performing fast contractions between tensors of arbitrary rank.

\section{Results}
\label{sec: validating 2pr}
We now present results, where the $\Delta \rm{NNLO_{GO}}(394)$ interaction~\cite{DeltaGo2020} at next-to-next-to-leading order in the chiral expansion is used. The latter has been applied successfully to several physics cases~\cite{Sun2025,Bonaiti2024,PbAbInitio}. 
All our computations start from a spherical HF state built from a model space consisting of up to $N_{max}$+1 =13 major harmonic oscillator (HO) shells.
Model-space convergence is controlled by $N_{max}$ and by the oscillator frequency $\hbar\omega$, which we typically vary between 10 and 18.
The optimal frequency for a given observable is determined as the one for which the discrepancy between predictions for $N_{max}=10$ and $N_{max}=12$ is smallest.
The deviation at the optimal frequency provides an estimate of the basis-truncation uncertainty. Three-nucleon force matrix elements are included up to $E_{3max} = 16$, which is deemed accurate up to the Ca chain~\cite{DeltaGo2020,Tichai2024}.

Ground-state calculations for closed-shell nuclei are performed at the CCSD and CCSDT-1 levels, which we denote for short as $D$ and $T$-1, respectively. In the latter case, an additional truncation is imposed on the size of the $3p$-$3h$ space by setting the energy cut $E_{pqr} = e_p + e_q + e_r < \Tilde{E}_{3\rm{max}}$, where $e_p = \abs{N_p - N_F}$ is the energy difference of the single-particle energies $N_p$ with respect to the Fermi surface $N_F$~\cite{DeltaGo2020}.
Throughout this work, we have used $\Tilde{E}_{3\rm{max}}=500$ MeV. This allows for the inclusion of most of the relevant $3p$-$3h$ configurations, while keeping the computational cost manageable.

\begin{table}[!ht]
\caption{Notation used for  closed-shell and 2PA/2PR calculations. Truncations employed for the ground (excited) states are on the left (right) of the symbol ``/", respectively.}
\label{tab: 2pr truncations}
\begin{ruledtabular}
\begin{tabular}{llll}
       & Ground state & EOM &  Truncation scheme \\  
        \noalign{\vskip 1.mm} 
        \hline
        \noalign{\vskip 1.5mm} & CCSD & & $D$ \\
        \noalign{\vskip 1.5mm} Closed-shell & CCSD & & $T$-1 \\
        \noalign{\vskip 1.5mm} & CCSD & CCSD & $D/D$ \\
        \hline
        \noalign{\vskip 1.5mm} 
            & CCSD & $2p$-$0h$ & $D/2p$-$0h$ \\
        \noalign{\vskip 1.5mm} 
        2PA & CCSD & $3p$-$1h$ & $D/3p$-$1h$ \\
        \noalign{\vskip 1.5mm} 
            & CCSDT-1 & $3p$-$1h$ & $T$-1$/3p$-$1h$ \\
        \hline
        \noalign{\vskip 1.5mm} 
            & CCSD & $0p$-$2h$ & $D/0p$-$2h$ \\
        \noalign{\vskip 1.5mm} 
        2PR & CCSD & $1p$-$3h$ & $D/1p$-$3h$ \\
       \noalign{\vskip 1.5mm} 
            & CCSDT-1 & $1p$-$3h$ & $T$-1$/1p$-$3h$ \\
\end{tabular}
\end{ruledtabular}
\end{table}

For EOM-CC, two truncations are involved, namely at the level of the g.s., i.e. the amplitudes $\hat{T}$ and $\hat{\Lambda}$, and that of the EOM excitation amplitudes $\hat{R}_f$, $\hat{L}_f$. 
The choice of the truncation in the g.s.~affects the EOM calculation through $\Bar{H}$ and the reference energy $\Delta E_{0}^{*}$.
The notation employed to denote different approximations is summarized in Tab.~\ref{tab: 2pr truncations} for the case of closed-shell and particle-attached/removed nuclei.
Shorthand notations for the schemes that are obtained by varying the g.s.~and EOM truncations are also introduced.

\subsection{Ground state energies}
\label{sec: gs energies}
We start by validating the 2PR-EOM approach on the binding energies of the $^{22}\rm{O}$ and $^{38}\rm{Ca}$ nuclei.
Our results are summarized in  Tab.~\ref{tab: energies 2pr}. 
The g.s.~energy predicted by a given truncation scheme is shown in the last column.
For 2PR (2PA) calculations, we also report the energy of the reference nucleus ($E_{0}^{*}$) and the excitation energy of the 2PR (2PA) system wrt. the reference ($\omega_0$).
\begin{table}[!ht]
\caption{Ground-state energies  of $^{22}\rm{O}$ and $^{38}\rm{Ca}$ and corresponding model-space uncertainties for different truncation schemes.
For 2PR/2PA-EOM calculations, the energy of the closed-shell reference $E_0^{*}$ and the excitation energy $\omega_0$ wrt. the reference are also reported.}
\label{tab: energies 2pr}
\begin{ruledtabular}
\begin{tabular}{ p{1.cm} p{2.cm} p{1.7cm} p{1.7cm} p{1.8cm} }
         & Truncation & $E_0^{*}$ [MeV] & $\omega_0$ [MeV] & $E_0$ [MeV]  \\  
        \noalign{\vskip 1.mm} 
        \hline
        \noalign{\vskip 1.5mm} 
         $^{22}\rm{O}$ 
         & Experiment~\cite{wang2021ame} & & & -162.0 
        \\ 
        \noalign{\vskip 2.mm} 
        \cline{2-5}
        \noalign{\vskip 2.mm} 
         & $D$ & & & -153.5 (1) \\
        \noalign{\vskip 1.5mm} 
                      & $T$-1 & & & -161.5 (1) \\
        \noalign{\vskip 2.mm} 
        \cline{2-5}
        \noalign{\vskip 2.mm} 
                      & $D/0p$-$2h$ & -160.5 & 10.6 & -149.8 (2) \\
        \noalign{\vskip 1.5mm} 
                      & $D/1p$-$3h$ & -160.5 & 6.7 & -153.8 (1) \\
        \noalign{\vskip 1.5mm} 
                      & $T$-1$/1p$-$3h$ & -169.5 & 8.3 & -161.2 (3) \\
        \noalign{\vskip 2.5mm} 
        \hline
        % Ca38
        \noalign{\vskip 2.5mm} 
        $^{38}\rm{Ca}$ 
        & Experiment~\cite{wang2021ame} & & & -313.1 
        \\
        \noalign{\vskip 2.mm} 
        \cline{2-5}
        \noalign{\vskip 2.mm} 
                       & $D/2p$-$0h$ & -266.7 & -12.6 & -279.3 (4) \\
        \noalign{\vskip 1.5mm} 
                       & $D/3p$-$1h$ & -266.7 & -28.3 & -295.0 (4) \\
        \noalign{\vskip 1.5mm} 
                       & $T$-1$/3p$-$1h$ & -284.2 & -26.88 & -311.1 (5) \\
        \noalign{\vskip 2.mm} 
        \cline{2-5}
        \noalign{\vskip 2.mm} 
                       & $D/0p$-$2h$ & -324.7 & 37.8 & -286.9 (4) \\
        \noalign{\vskip 1.5mm} 
                       & $D/1p$-$3h$ & -324.7 & 28.2 & -296.5 (4)\\
        \noalign{\vskip 1.5mm} 
        % hw=18
                       & $T$-1$/1p$-$3h$ & -346.0 & 31.38 & -314.6 (4)
    \end{tabular}
\end{ruledtabular}
\end{table}

$^{22}\rm{O}$ can be obtained by removing two neutrons from $^{24}\rm{O}$, the most neutron-rich bound oxygen isotope. 
Strong experimental evidence, supported by theoretical calculations~\cite{Hagen2012Oxygen,Otsuka2010}, suggests that both $^{22}\rm{O}$~\cite{THIROLF200016} and $^{24}\rm{O}$~\cite{KanungoO24,NeupaneO24} are spherical closed-shell nuclei.
The $D/0p$-$2h$ scheme represents the simplest 2PR approximation and recovers approximately  85\% of the experimental binding energy (from Ref.~\cite{wang2021ame}).
The inclusion of $1p$-$3h$ excitations significantly improves the description, contributing an additional correlation energy of about 8 MeV.
The resulting $D/1p$-$3h$ truncation compares closely with closed-shell CCSD energies, as also shown in~\cite{Marino2024Nsd}.
Furthermore, we introduce a new scheme, dubbed $T$-1$/1p$-$3h$, which refines the description of the reference nucleus by performing a $T$-1 g.s.~calculation. 
By including leading-order triple contributions in the g.s.~equations, this approach improves the predictive power considerably, leading to a better agreement with experiment for both the closed-shell and 2PR frameworks. 
This substantial improvement in predicting the binding energy of the open-shell isotope arises from a balance between two factors, namely, the gain in the energy $E_0^{*}$ of the $^{24}\rm{O}$ reference nucleus, amounting to approximately -9 MeV, and a positive correction of about 1.5 MeV to the 2PR-EOM excitation energy $\omega_0$ of $^{22}\rm{O}$.
Overall, the $T$-1$/3p$-$1h$ prediction is lower than the $D/1p$-$3h$ one by about 7.5 MeV and close to data. 
\begin{figure}[h]
    \centering
    \includegraphics[width=\columnwidth]{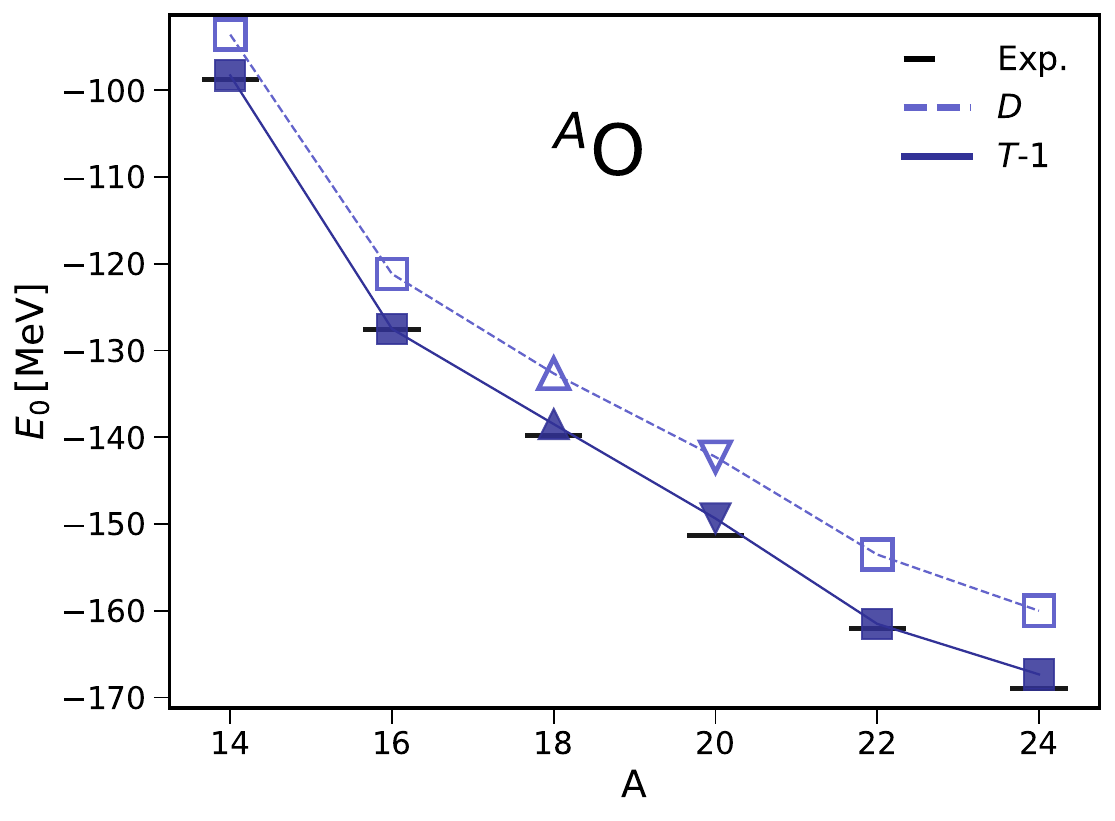}
    \includegraphics[width=\columnwidth]{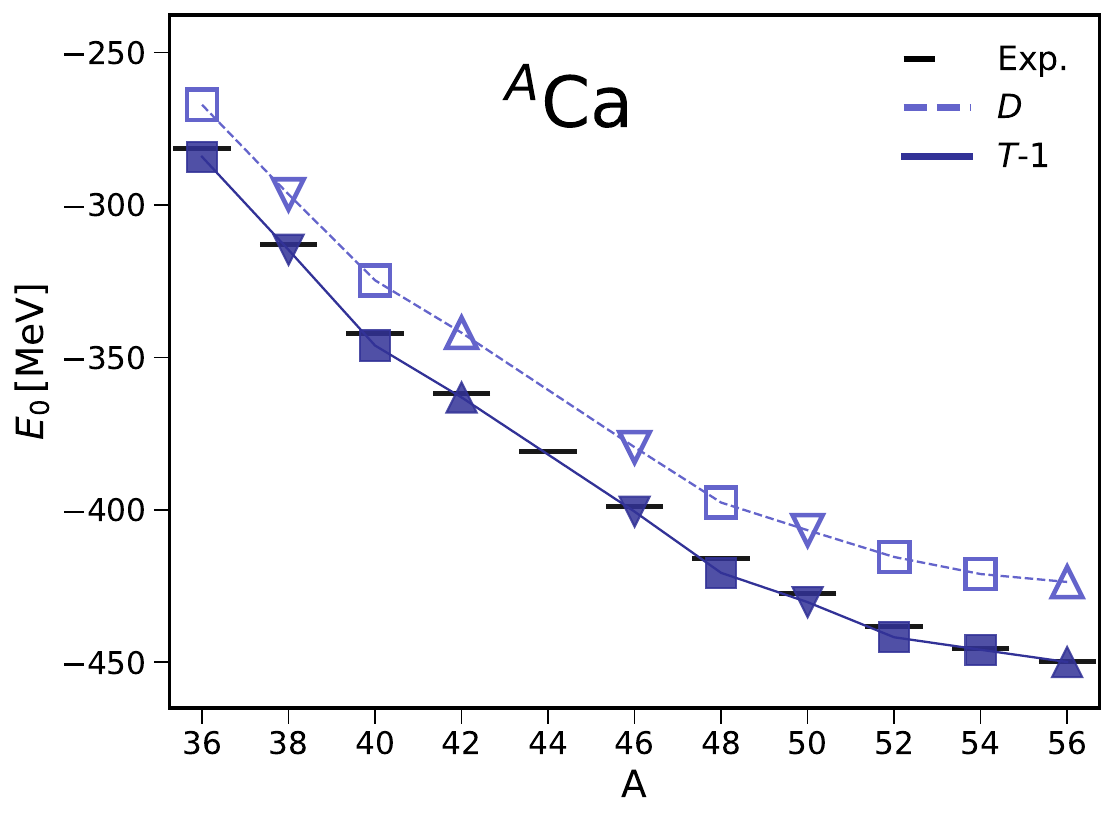}
    \caption{Binding energies of $\rm{O}$ (top) and $\rm{Ca}$ (bottom) isotopes.
    Empty (filled) markers denote computations where the closed-shell g.s.~is described at the CCSD (CCSDT-1) truncation level, labeled as $D$ ($T$-1).
    Calculations performed with closed-shell CC (squares), 2PA-EOM (upward arrows), and 2PR-EOM (downward arrows) using the $\Delta \rm{NNLO_{\mathrm{GO}}(394)}$ interaction are compared to experimental data from Ref.~\cite{wang2021ame}.
    Lines are a guide to the eye.
    }
    \label{fig: binding energies chains}
\end{figure}

Next, we examine the $^{38}\rm{Ca}$ isotope, which can be computed by either removing two neutrons from $^{40}\rm{Ca}$ using the 2PR ansatz, or adding two neutrons to $^{36}\rm{Ca}$ within the 2PA scheme~\cite{Jansen2013}.
This makes $^{38}\rm{Ca}$ an ideal case for testing the consistency of the two EOM ans\"{a}tze.
As before, binding energy calculations are performed at different truncation levels for both the g.s.~and the EOM diagrams (see Tab.~\ref{tab: energies 2pr}).
When considering only $2p$-$0h$ excitations for 2PA and $0p$-$2h$ excitations for 2PR, the difference in binding energy amounts to several MeV. However, when 1$p$-$3h$ contributions are included for 2PR and $3p$-$1h$ ones for 2PA, the two approaches converge to similar g.s.~energies.
As observed before, the inclusion of triples contributions in the reference state recovers the vast majority of the correlation energy, allowing to reach a good agreement with the experiment.
The remaining discrepancy between 2PA and 2PR amounts to just 1\% of the total g.s.~energy of $^{38}\rm{Ca}$.
The primary impact of $T$-1 amplitudes is a lowering of the energy of the $^{40}\rm{Ca}$ ($^{36}\rm{Ca}$) reference of approximately 20 MeV.
Meanwhile, excitation energies $\omega_0$ receive a positive correction of 1.5 and 3 MeV in 2PA and 2PR, respectively, equivalent to ~5-10\% of their value.

Finally, binding energies across the $\rm{O}$ and $\rm{Ca}$ chains are presented in the top and bottom panels of Fig.~\ref{fig: binding energies chains}, respectively.
For even-mass nuclei, two sets of calculations are reported in each panel.
Empty symbols denote computations where the nucleus itself (for closed-shell CC) or the closed-shell reference $\ket{\Psi_0}$ (for open-shell isotopes) is determined at the CCSD level ($D$ in the legend), while filled markers indicate calculations where the g.s.~is treated with CCSDT-1 (denoted as $T$-1). 
By combining the EOM approaches, a large number of isotopes in semi-magic isotopic chains can be addressed, and a fairly comprehensive picture of their binding energies can be obtained.
Overall, with the inclusion of triples excitations in the g.s., the results show good agreement with experimental data, not only for the well-established closed-shell CC calculations but also for the particle-attached/removed techniques.

\subsection{Excited states}
\label{sec: excited states}

\begin{figure*}
    \centering
    \includegraphics[width=1\textwidth]{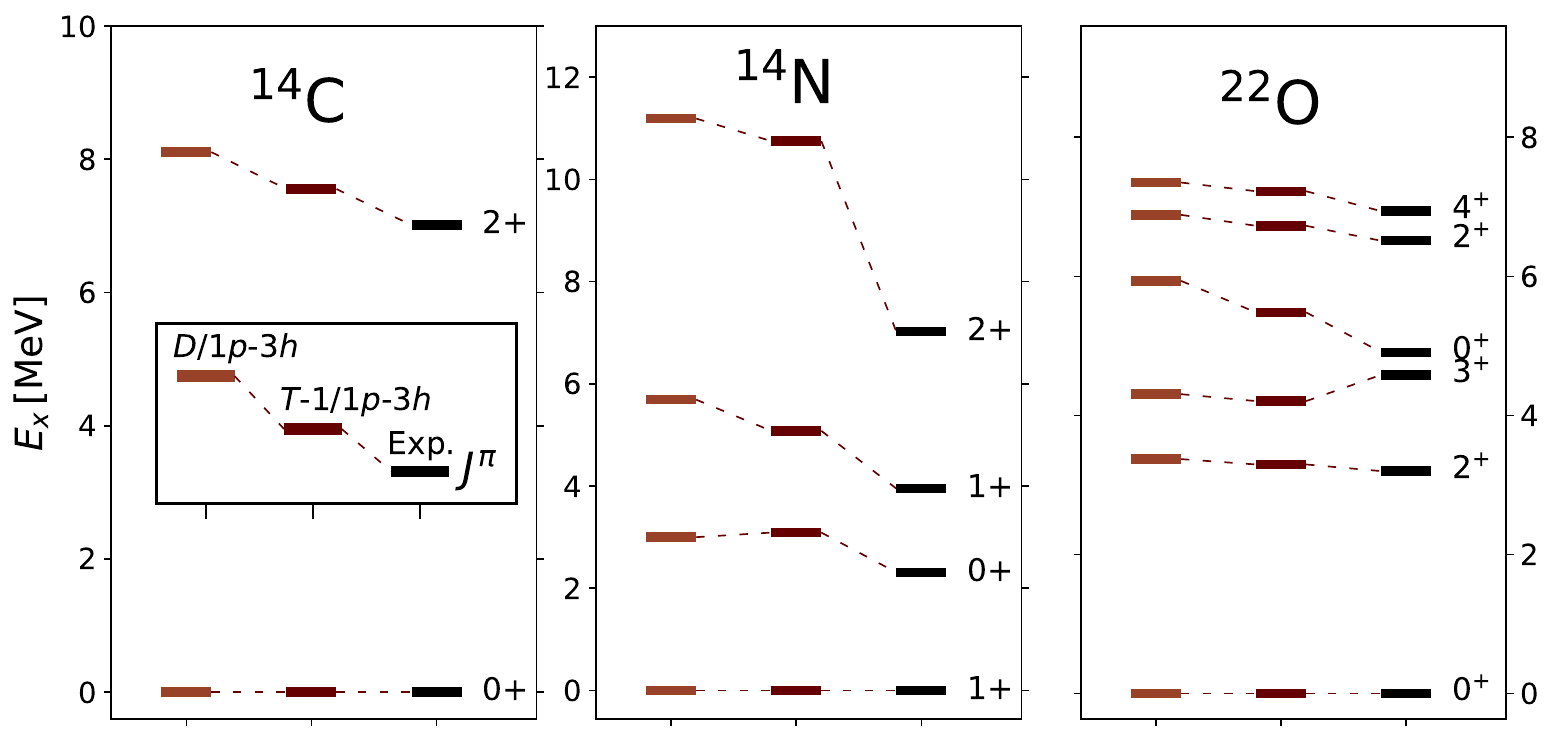}
    \caption{
    Excitation energies (in MeV) of selected states computed for several nuclei using the 2PR-EOM approach with the $\Delta \rm{NNLO_{GO}}(394)$ interaction.
    The predictions of the $D$$/3p$-$1h$ and $T$-$1/3p$-$1h$ truncation schemes are compared to experimental data from Refs.~\cite{DataA14,DataA22}.
    }
    \label{fig: Spectrum_2PR}
\end{figure*}

In this Section, we investigate selected excited states of 2PR nuclei in the oxygen region.
Our main results are shown in Fig.~\ref{fig: Spectrum_2PR}, where excitation energies computed with the $D/1p$-$3h$ and $T$-$1/1p$-$3h$ truncations are compared to experimental data from Refs.~\cite{DataA14,DataA22}.
In addition to 2PR-EOM predictions, in Tab.~\ref{tab: exc states compact} we report the results of closed-shell EOM-CC using the $D/D$ approximation.
Also, partial norms (see App.~\ref{sec: Partial norms}), which quantify the contributions of $0p$-$2h$ ($1p$-$3h$) excitations to the wave function in 2PR (closed-shell) EOM, are listed.

Using $^{16}\rm{O}$ as a reference, three nuclei can be described within the 2PR ansatz. By removing two protons or two neutrons, one obtains the closed-subshell isotopes $^{14}\rm{C}$ and $^{14}\rm{O}$, respectively. These two systems are mirror nuclei with similar structures; in the following, we focus on $^{14}\rm{C}$. 
Alternatively, removing one neutron and one proton gives access to $^{14}\rm{N}$, which can also be studied using the closed-shell EOM approach via a charge-exchange calculation based on either $^{14}\rm{O}$ or $^{14}\rm{C}$~\cite{Ekstrom2014}.

\begin{table*}
    \caption{
    Excitation energies (in MeV) of selected nuclear states with given angular momentum and parity $J^{\pi}$. Experimental values~\cite{DataA14,DataA22} are compared to calculations performed with the  closed-shell EOM $D/D$ truncation and two 2PR-EOM approximations ($D/1p$-$3h$, $T$-$1/1p$-$3h$). 
    Partial norms $n(1p$-$1h)$ [$n(0p$-$2h)$] are also reported for the closed-shell ($D/1p$-$3h$) computations.
    }
    \label{tab: exc states compact}
    \begin{ruledtabular}
    \begin{tabular}{ p{1.2cm} p{1.2cm} p{1.4cm} | p{1.7cm} p{1.7cm} | p{1.7cm} p{1.7cm} p{1.7cm}  }
        & & & Closed-shell & & & 2PR-EOM & \\
        \rule{0pt}{3.5mm}
                         & $J^{\pi}$ & Exp. & $D/D$ & $n(1p$-$1h)$ &
        $D/1p$-$3h$ & $T$-$1/1p$-$3h$ & $n(0p$-$2h)$ \\  
        \hline
        \rule{0pt}{3.5mm}
                         & $1^{-}$ &  6.09 & 9.15 &  0.86 & 12.83 & 13.17 & 0.19 \\
          $^{14}\rm{C}$  & $2^{+}$ &  7.01 & 7.67 &  0.88 & 8.11  & 7.55  & 0.89 \\
        \hline
        \rule{0pt}{3.5mm}
                      & $1^{-}$ & 5.17 & 8.65 &  0.87 & 12.96 & 13.29 & 0.19 \\
        $^{14}\rm{O}$ & $2^{+}$ & 6.59 & 7.65 &  0.87 & 8.20  & 7.66  & 0.89 \\
        \hline
        \rule{0pt}{3.5mm}
                      & $0^{+}$ & 2.31 & 3.13 & 0.88 & 3.00 & 3.09  & 0.87 \\
        $^{14}\rm{N}$ & $1^{+}$ & 3.95 & 9.09 & 0.83 & 5.69 & 5.08  & 0.91 \\
                      & $2^{+}$ & 7.02 & 8.55 & 0.86 &11.19 & 10.75 & 0.89  \\
        \hline
        \rule{0pt}{3.5mm}
                      & $2^{+}$ & 3.20 & 3.35 & 0.92 & 3.37 & 3.29 & 0.93 \\
                      & $3^{+}$ & 4.58 & 4.45 & 0.91 & 4.31 & 4.21 & 0.93 \\
        $^{22}\rm{O}$ & $0^{+}$ & 4.91 & 11.26& 0.52 & 5.93 & 5.48 & 0.92 \\
                      & $2^{+}$ & 6.51 & 7.79 & 0.91 & 6.88 & 6.73 & 0.93 \\
                      & $4^{+}$ & 6.94 & 7.30 & 0.92 & 7.35 & 7.22 & 0.93 \\
    \end{tabular}
    \end{ruledtabular}
\end{table*}

$^{14}\rm{C}$ exhibits a $2^{+}$ state at a relatively high excitation energy.
In a naive shell model picture, this state is predominantly characterized by a $1p$-$1h$ configuration, where a proton is promoted from the occupied $p_{\frac{3}{2}}$ orbital to the empty $p_{\frac{1}{2}}$ orbital.
Similarly, within the 2PR framework, the leading-order description of the $2^{+}$ state arises from the removal of two protons from the $p$ shell of $^{16}\rm{O}$.
Its relatively simple structure makes it a good target for the EOM ansatz.
Indeed, a reasonable description is achieved with the closed-shell $D$/$D$ truncation, which overestimates the experimental excitation energy by approximately 600 keV.
While 2PR-EOM at the $D$/$1p$-$3h$ truncation level slightly surpasses the closed-shell EOM prediction, the $T$-$1/1p$-$3h$ approximation closely aligns with the $D$/$D$ result.

Negative-parity states are a difficult testbed for EOM, as they are predominantly governed by cross-shell configurations ~\cite{Jansen2013}.
The $1^{-}$ state of $^{14}\rm{C}$ is examined as a prototypical case.
The only $2h$ configurations that yield negative parity arise from the removal of one proton from the $s$ shell and another from either the $p_{\frac{1}{2}}$ or $p_{\frac{3}{2}}$ orbital in the $^{16}\rm{O}$ reference.
Similarly, in the closed-shell EOM approach, particle-hole excitations contributing to this state involve promoting a proton from the $p$ shell to the opposite-parity $sd$ major shell. 
Quantitatively, closed-shell EOM provides more accurate results, whereas 2PR-EOM overestimates the $1^{-}$ energy by several MeV. 
The inaccurate description of this state at the level of $1p$-$3h$ excitations is correlated to a relatively small $n(0p$-$2h)$ norm, as low as 20\%, indicating that significant higher-order contributions are missing.

The odd-odd $^{14}\rm{N}$ nucleus exhibits a different structure, characterized by a $1^{+}$ g.s.
We first predict the low-lying positive-parity excited states using charge-exchange EOM~\cite{Ekstrom2014}.
The results are satisfactory for the $0^{+}$ and $2^{+}$ states,
however, the excitation energy of the second $1^{+}$ state is overestimated, placing it above the $2^{+}$ state and exceeding the experimental value by several MeV.
A similar trend was observed in Ref.~\cite{NNLOsat} when applying the same method with the $\rm{NNLO_{sat}}$ interaction.
Notably, the 2PR-EOM approach successfully reproduces the correct level ordering, as shown in Fig.~\ref{fig: Spectrum_2PR}, already at the $D$/$1p$-$3h$ truncation level.
The inclusion of ground-state triples further lowers the excitation energies by less than 600 keV, bringing theoretical predictions into closer agreement with experimental data. 

Finally, the most intriguing results are obtained for $^{22}\rm{O}$.
2PR-EOM accurately reproduces the correct ordering of the five lowest-lying excited states and outperforms the conventional closed-shell ansatz when compared with data.
The results of the $T$-$1/1p$-$3h$ truncation differ from the $D$/$1p$-$3h$ predictions by roughly 100 keV, except for the second $0^{+}$ state, where the discrepancy is around 0.5 MeV.
In general, these results are in good agreement with the experimental data. The relatively mild sensitivity of excitation energies to the inclusion of g.s. triples suggests that our predictions are robust and subject to a small many-body truncation error.

The accuracy of the 2PR-EOM ansatz in the case of $^{22}$O can be attributed to the underlying level structure of this nucleus. 
Within the shell model picture, $^{22}\rm{O}$ and $^{24}\rm{O}$ are closed-subshell systems, with valence neutrons occupying the $1d_{\frac{5}{2}}$ orbital in $^{22}\rm{O}$ and both the $1d_{\frac{5}{2}}$ and $2s_{\frac{1}{2}}$ orbitals in $^{24}\rm{O}$~\cite{Brown2005,KanungoO24,Otsuka2010}.
These orbitals lie close in energy to each other and to the unoccupied $1d_{\frac{3}{2}}$ orbital, facilitating the formation of multiple positive-parity excited states via intra-shell excitations~\cite{DeGregorio2018}. This pattern is reflected in the experimentally observed spectrum~\cite{DataA22}. 
All the states reported in Tab.~\ref{tab: exc states compact} can be interpreted as arising from the coupling of two nucleons within the $sd$ shell.
This interpretation is further supported by consistently large $1p$-$1h$ and $0p$-$2h$ partial norms, which exceed 0.9 in all cases except for the second $0^{+}$ state in closed-shell EOM. For this state, the excitation energy is also significantly overestimated.

\subsection{Electric dipole response}
\label{sec: valid 2pr response}
We now move on to the study of the nuclear response functions and their sum rules, focusing on the case of the electric dipole operator. The latter is defined as 
\begin{align}
    \hat{\Theta} = \sum_{i=1}^{A}
    ( \hat{\mathbf{r}}_i - \hat{\mathbf{R}}_{cm} )
    \left(
    \frac{1 + \hat{\tau}_i^z }{ 2 }
    \right),
\end{align}
where $\hat{\mathbf{r}}_i$ and $\hat{\mathbf{R}}_{cm}$ are the positions of the $i$-th particle and the center of mass, respectively, and $\hat{\tau}_i^z$ is the isospin projection.
The open-shell nuclei we consider have angular momentum $J^{\pi} = 0^{+}$ in their g.s., in which case, the electric dipole operator only allows for transitions to the $1^{-}$ excited states~\cite{Bonaiti2024}.

The electric dipole polarizability $\alpha_D$ is defined as the inverse energy-weighted sum rule  of the dipole response function as
\begin{align}
    \label{eq: alphaD}
    \alpha_D = 2\alpha \int d\omega \frac{R(\omega)}{\omega} 
    = 2\alpha \lim_{\Gamma \to 0} \int d\sigma \frac{L(\sigma,\Gamma)}{\sigma},
\end{align}
where $\alpha$ is the fine structure constant.
In this paper, $\alpha_D$ is obtained by computing the LIT with a small width $\Gamma=10^{-4}\,\rm{MeV}$ and by integrating it numerically as in the r.h.s.~of Eq.~\eqref{eq: alphaD}.

While evaluating the LIT involves two steps for closed-shell nuclei~\cite{Bacca2014,Miorelli2016}, it takes three in open-shell systems. First, one determines the g.s.~for the closed-shell reference. Second, the amplitudes $\hat{R}_{0}$ and $\hat{L}_{0}$ and the excitation energy $\omega_0$ for the g.s.~of the open-shell system are found by solving the EOM equations. 
Third, the auxiliary states~\eqref{eq: aux state 2pr lit right} and~\eqref{eq: aux state 2pr lit left} are determined by means of the Lanczos method~\cite{Bonaiti2024,Bacca2014}.  
For the last step, we use 60 Lanczos coefficients, which are sufficient for computing $\alpha_D$.
Results are reported for a model space with $N_{max}=12$ at the optimal HO frequency.
The similarity-transformed operators $\Bar{\Theta}$ and $\overline{ \Theta^{\dagger} }$ are evaluated at the CCSD level, as in Eq.~(27) of Ref.~\cite{Miorelli2018}.
In the case of closed-shell nuclei, we vary the truncation level used in the ground state between CCSD and CCSDT-1, while we solve the EOM equations for the auxiliary states including contributions up to $2p$-$2h$ excitations.
We will use the compact notation $D$ and $T$-1 to refer to these truncations for the LIT-CC approach.
In the case of 2PR and 2PA nuclei, we vary both the truncation employed in the closed-shell g.s.~and in the EOM amplitudes $\hat{R}_0$ and $\hat{L}_0$.
The auxiliary states~\eqref{eq: aux state 2pr lit right} and \eqref{eq: aux state 2pr lit left} are always evaluated including up to $1p$-$3h$ ($3p$-$1h$) excitations in the pivots and the matrix-vector product.
We keep track of all these different truncations by using the notation introduced in Tab.~\ref{tab: 2pr truncations}.

We first test the predictions of $\alpha_D$ for $^{22}\rm{O}$ and $^{38}\rm{Ca}$:  we compare the 2PR and closed-shell CC variants for the closed subshell nucleus of $^{22}\rm{O}$,  while for  $^{38}\rm{Ca}$ we compare the 2PR and 2PA schemes. 
Predictions for different truncations and their model-space uncertainties are reported in Tab.~\ref{tab: alphaD 2pr}.
\begin{table}[!ht]
\begin{ruledtabular}
\begin{tabular}{ p{1.cm} p{2.5cm} p{2.5cm} }
         & Truncation & $\alpha_D\,[\rm{fm}^{3}]$ \\  
        \noalign{\vskip 1.mm} 
        \hline
        \noalign{\vskip 2.mm} 
        $^{22}\rm{O}$  
        & $D$ & 0.859 (4) \\
        \noalign{\vskip 1.5mm} 
        & $T$-1 & 0.824 (3) \\
        \noalign{\vskip 2.mm} 
        \cline{2-3}
        \noalign{\vskip 2.mm} 
        & $D/0p$-$2h$ & 0.73 (1) \\
        \noalign{\vskip 1.5mm} 
        & $D/1p$-$3h$ & 0.58 (1)\\
        \noalign{\vskip 1.5mm} 
        & $T$-1$/1p$-$3h$ & 0.55 (1) \\
        \noalign{\vskip 2.5mm} 
        \hline
        \noalign{\vskip 2.mm} 
        $^{38}\rm{Ca}$ 
        & $D/2p$-$0h$ & 1.894 (6) \\
        \noalign{\vskip 1.5mm} 
        & $D/3p$-$1h$ & 1.083 (1) \\
        \noalign{\vskip 1.5mm} 
        & $T$-1$/3p$-$1h$ &  1.03 (1) \\
        \noalign{\vskip 2.mm} 
        \cline{2-3}
        \noalign{\vskip 2.mm} 
        & $D/0p$-$2h$ & 1.63 (4) \\
        \noalign{\vskip 1.5mm} 
        & $D/1p$-$3h$ & 1.039 (2)\\
        \noalign{\vskip 1.5mm} 
        & $T$-1$/1p$-$3h$ & 0.961 (7)
    \end{tabular}
\end{ruledtabular}
\caption{Electric dipole polarizability $\alpha_D$ of $^{22}\rm{O}$ and $^{38}\rm{Ca}$ and corresponding model-space uncertainties for different truncation schemes.}
\label{tab: alphaD 2pr}
\end{table}

For $^{22}\rm{O}$, closed-shell LIT-CC results show good agreement with the experimental value $\alpha_D = 0.87\,(9)\,\rm{fm}^{3}$ (see~\cite{O22response,Miorelli2016}), though the latter has significant uncertainties.
G.s.~triples ($T$-1) have a relatively modest effect, leading to a slight reduction in $\alpha_D$ compared to the $D$ case.
In contrast, 2PR-LIT predictions tend to underestimate $\alpha_D$ relative to closed-shell CC. The dominant source of variation is the EOM truncation rather than the g.s.~approximation, as the difference between $D/0p$-$2h$ and $D/1p$-$3h$  is much larger than that between $D/1p$-$3h$ and $T$-1$/1p$-$3h$. 
This trend is consistent with the closed-shell case and suggests that many-body errors should be estimated by comparing different EOM approximations, as done for the 2PA case~\cite{Bonaiti2024}.

For the 2PR and 2PA ansatz applied to $^{38}\rm{Ca}$, the $D/1p$-$3h$ and $D/3p$-$1h$ schemes yield similar $\alpha_D$ values, with 2PR results being approximately 5\% lower than those of 2PA. The inclusion of triples in the reference state has only a minor effect, modifying $\alpha_D$ by a few percent, which is less significant than the difference observed between the $D/0p$-$2h$ and $D/1p$-$3h$ truncations.

\begin{figure*}[t]
    \centering
    \includegraphics[width=0.45\textwidth]{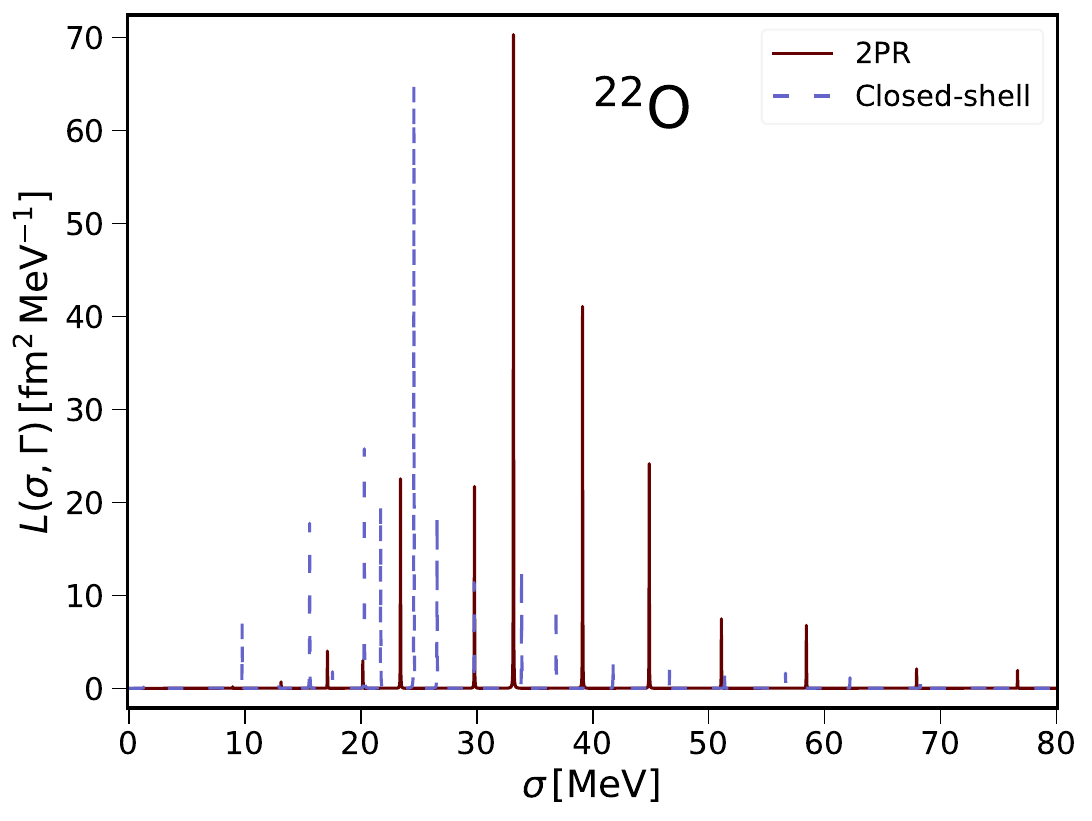}
    \includegraphics[width=0.45\textwidth]{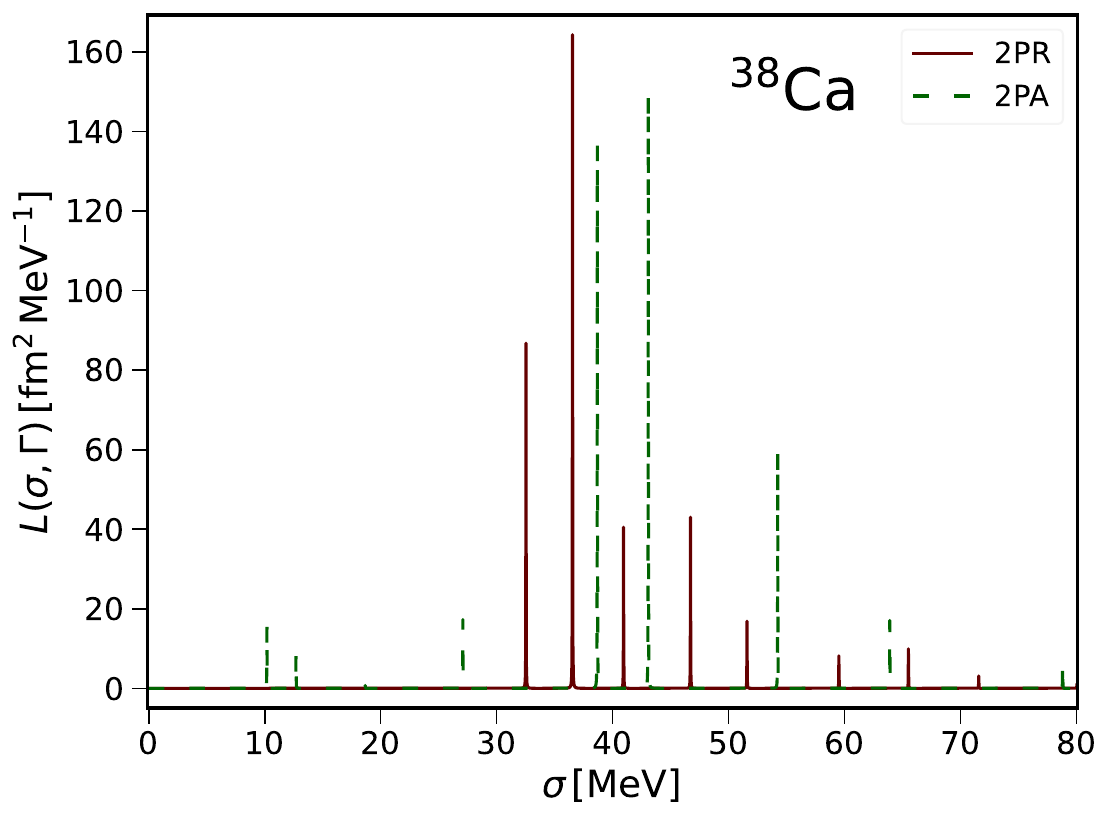}
    \includegraphics[width=0.45\textwidth]{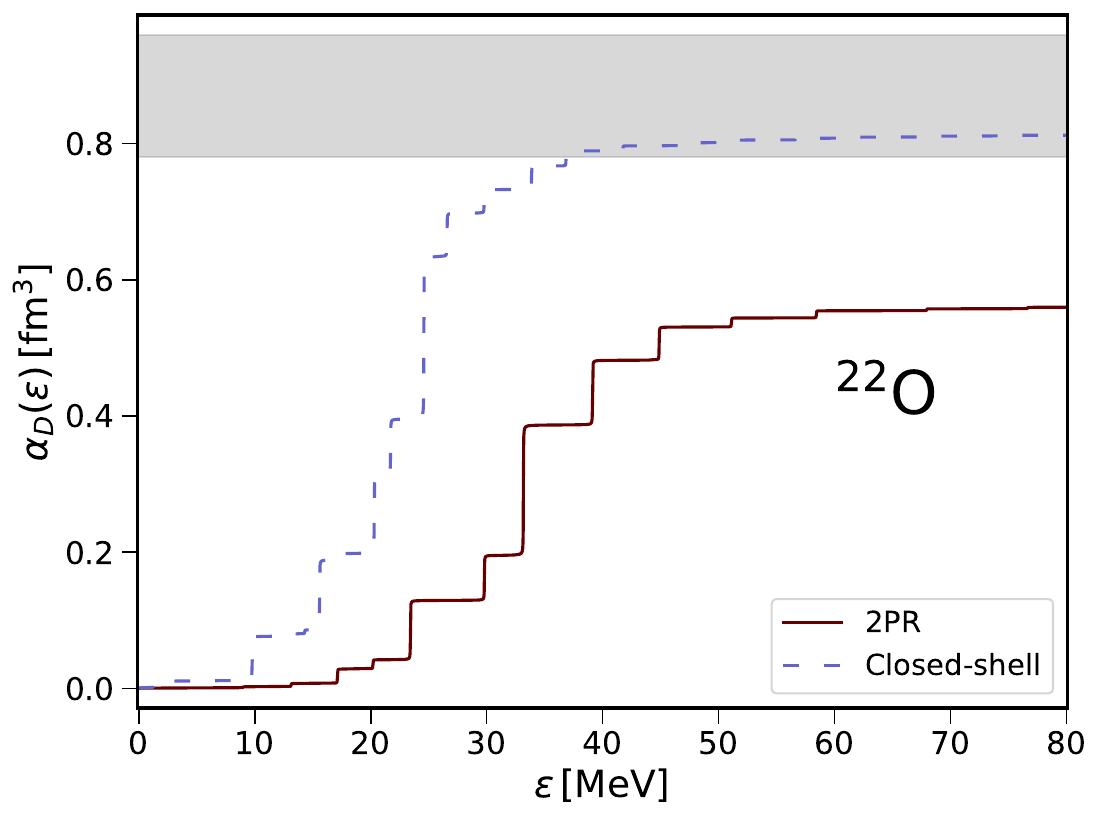}
    \includegraphics[width=0.45\textwidth]{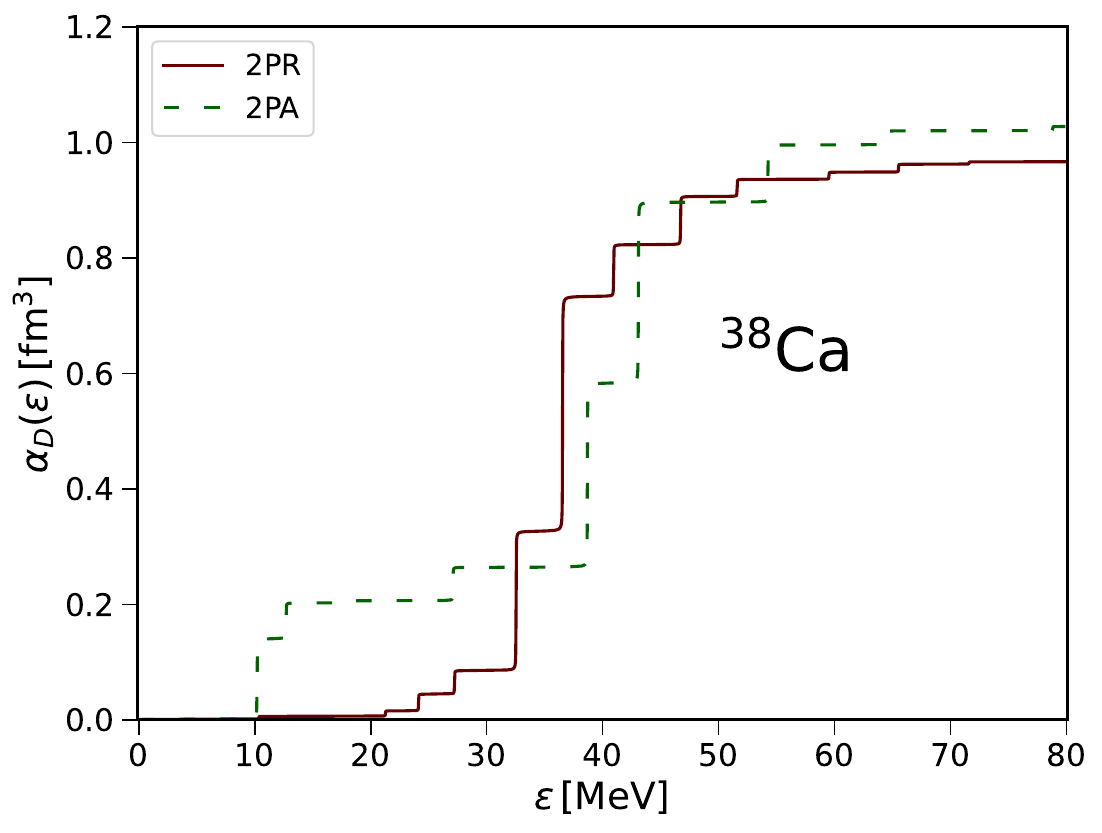}
    \caption{(Top) Discretized response with $\Gamma = 0.01\,\rm{MeV}$ for the $^{22}\rm{O}$ (left) and $^{38}\rm{Ca}$ (right) nuclei. 2PR calculations at the $T$-1$/1p$-$3h$ level are compared to closed-shell (2PA) responses based on the $T$-1 ($T$-1$/3p$-$1h$) truncations for $^{22}\rm{O}$ ($^{38}\rm{Ca}$). For $^{22}\rm{O}$, the experimental value for $\alpha_D$ is shown as a grey horizontal band.
    (Bottom) Corresponding $\alpha_D(\epsilon)$ running sum for $^{14}\rm{O}$ (left) and $^{38}\rm{Ca}$ (right). See text for details. }
    \label{fig: Cfr_Lit_panels}
\end{figure*}

We further investigate the behavior of $\alpha_D$ by considering the discretized response functions for $^{22}\rm{O}$ (left) and $^{38}\rm{Ca}$ (right) in Fig.~\ref{fig: Cfr_Lit_panels} (top panel).
They are obtained by evaluating $L(\sigma,\Gamma)$ with $\Gamma = 0.01\,\rm{MeV}$ and represent a discretization of the response function $R(\omega)$, where each peak corresponds to an eigenvalue of the Lanczos-reduced  Hamiltonian~\cite{Bacca2014}.
For $^{22}\rm{O}$, the $T$-1 closed-shell and $T$-1$/1p$-$3h$ 2PR calculations are shown, while
for $^{38}\rm{Ca}$, we compare the $T$-1$/1p$-$3h$ and $T$-1$/3p$-$1h$ truncations for 2PR and 2PA, respectively.
The corresponding  running sum rules, defined as the integral functions
\begin{align}
    \alpha_D(\epsilon) = 2\alpha \int_{0}^{\epsilon} d\sigma 
    \frac{L(\sigma,\Gamma=10^{-4})}{\sigma}\,,
\end{align}
are reported in the lower panels of Fig.~\ref{fig: Cfr_Lit_panels}.
For large $\epsilon$, the total $\alpha_D$ value is recovered.

In $^{22}\rm{O}$, the underestimation of $\alpha_D$ compared to closed-shell LIT-CC stems from the dipole response being shifted to higher energies by the 2PR method. 
The dominant peak in the closed-shell calculation, and, in general, the strength in the giant dipole resonance region, is displaced by about 10 MeV in 2PR-LIT, leading to missing contributions in the total sum rule.
Additionally, the closed-shell CC response function shows a small peak at around 10 MeV, which might be consistent with a soft dipole mode observed experimentally~\cite{O22response} and in SCGF calculations~\cite{Raimondi2019}.
This soft mode is responsible for an enhancement of the electric dipole polarizability in neutron-rich isotopes~\cite{Aumann_2013,Raimondi2019}.

In contrast, 2PR-LIT shows little to no strength below 15 MeV and exhibits increased fragmentation at higher energies.
As seen in Fig.~\ref{fig: Cfr_Lit_panels} (bottom left panel), the closed-shell running sum rises steeply around 10 MeV, whereas in 2PR calculations, significant contributions to $\alpha_D(\epsilon)$ appear only near 20 MeV.
Moreover, we have observed that the $D/1p$-$3h$ and $T$-1$/1p$-$3h$ discretized response functions are nearly identical, with the latter displaying peaks at slightly higher energies, leading to a small decrease in $\alpha_D$.

The challenges in describing the dipole response mirror those encountered for the $1^{-}$ excited states in Sec.~\ref{sec: excited states}.
The $1^{-}$ Lanczos pseudostates exhibit small  $0p$-$2h$ partial norms, often below 0.2, suggesting that higher-order contributions in the 2PR (and 2PA~\cite{Bonaiti2024}) amplitudes may be necessary for an accurate reproduction of response functions.

In $^{38}\rm{Ca}$, the discretized response shows a concentration of strength between approximately 35 and 45 MeV, with two prominent peaks in the 2PA case and three in the 2PR case.
The 2PA response exhibits greater fragmentation in the low-energy region and around 50 MeV, while the 2PR LIT features an additional peak at 30 MeV.
Examining $\alpha_D(\epsilon)$, we observe variations in the low-energy strength distribution. However, these differences are compensated by the bulk of the dipole strength being concentrated in the 35–45 MeV range, leading to similar $\alpha_D$ estimates at sufficiently high energies.

\begin{figure}
    \centering
    \includegraphics[width=\columnwidth]{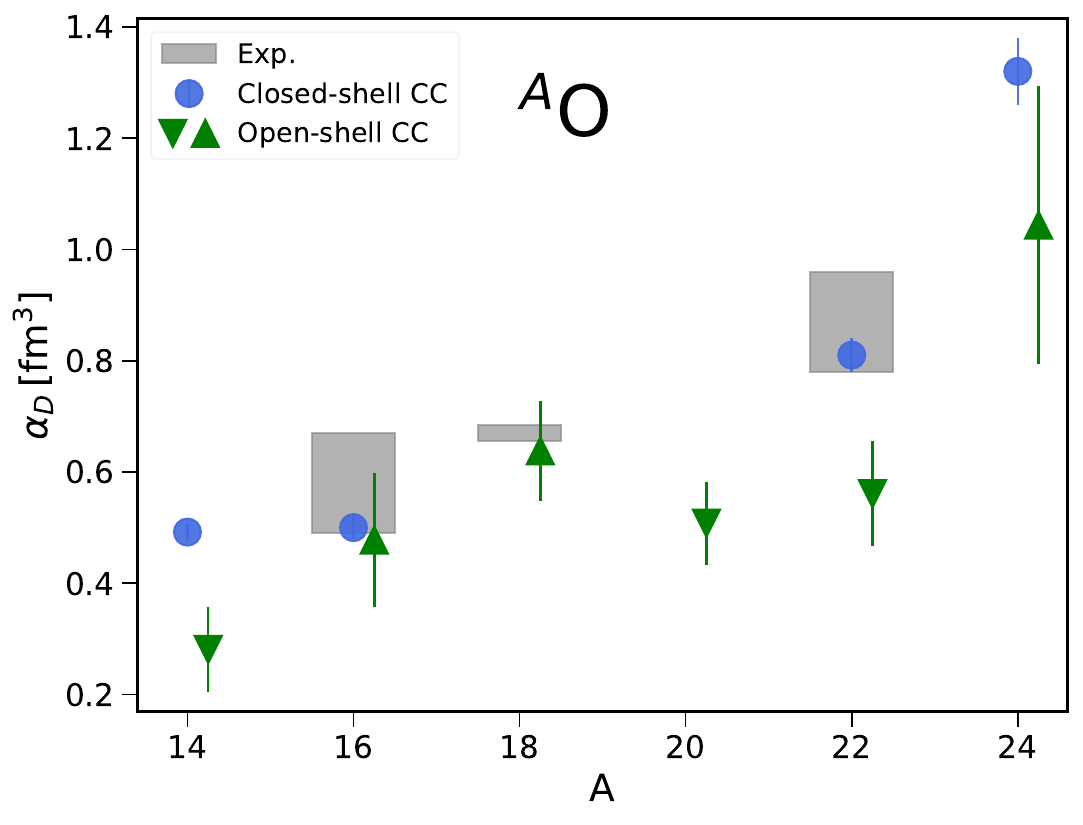}
    \includegraphics[width=\columnwidth]{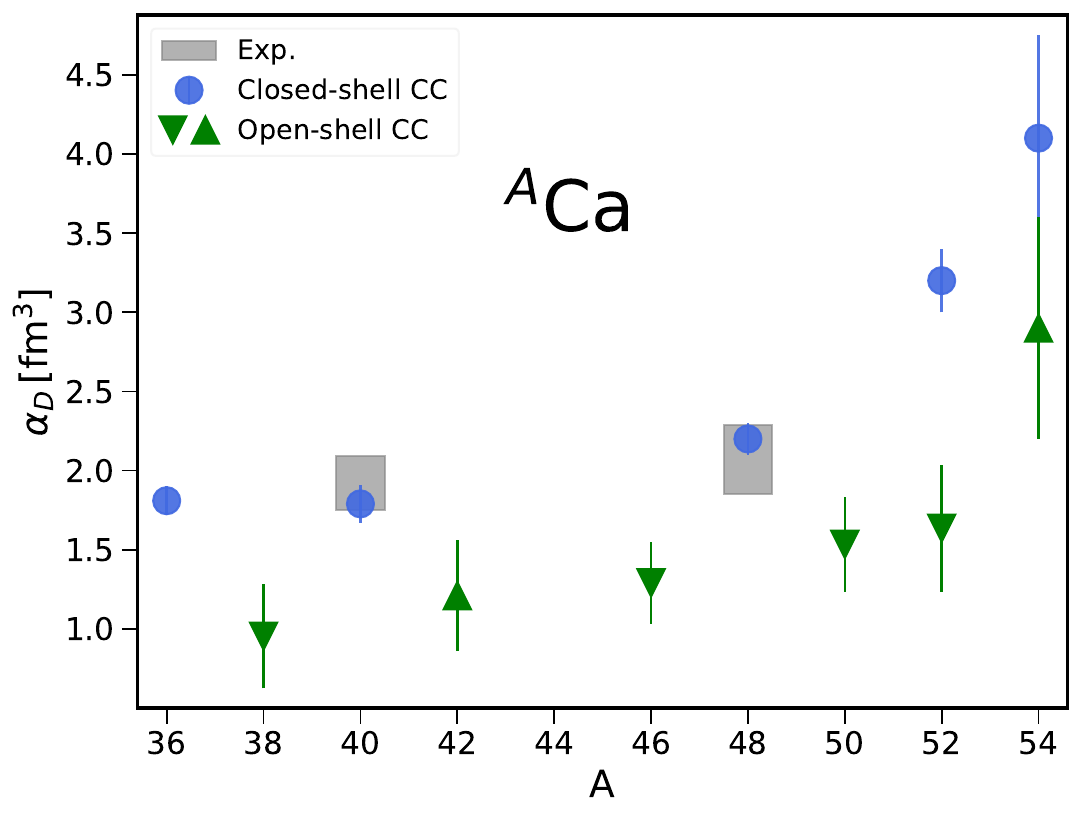}
    \caption{Electric dipole polarizability in the oxygen (top) and calcium (bottom) isotopic chains.
    Predictions of closed-shell LIT-CC (circles), 2PA-LIT-CC (upward triangles), and 2PR-LIT-CC (downward triangles) obtained using the $\Delta \rm{NNLO_{GO}}(394)$ interaction are compared to available experimental data.
    Error bars account for the model-space and many-body truncation errors.
    See text for details.
    }
    \label{fig: alphad chains}
\end{figure}

Finally, Fig.~\ref{fig: alphad chains} presents the evolution of $\alpha_D$ across the oxygen (top) and calcium (bottom) isotopic chains, combining results from LIT-CC approaches. 
For closed-shell nuclei, $T$-1 truncation results are shown as circles, while upward and downward triangles represent 2PA and 2PR calculations using the $T$-1$/3p$-$1h$ and $T$-1$/1p$-$3h$ approximations, respectively.
Theoretical uncertainties account for model-space and many-body truncation errors, while the Hamiltonian dependence is not addressed here. 
In closed-shell CC calculations, truncation uncertainty is estimated following Refs.~\cite{Miorelli2018,Bonaiti2024} as half the difference between the $D$ and $T$-1 calculations.
In open-shell nuclei, the EOM truncation represents the dominant source of uncertainty. We estimate the latter by comparing results at the $D/2p$-$0h$ and $T$-1$/3p$-$1h$ approximation level for 2PA systems, and at the $D/0p$-$2h$ and $T$-1$/1p$-$3h$ levels for 2PR ones.
Experimental data are also included for $^{16}\rm{O}$~\cite{AhrensO16}, $^{18}\rm{O}$~\cite{IshkhanovO18}, and $^{22}\rm{O}$~\cite{O22response} (see also~\cite{Bonaiti2024,Miorelli2016}).
In the calcium chain, available data are limited to the closed-shell isotopes  $^{40}\rm{Ca}$~\cite{BirkhanCa40} and $^{48}\rm{Ca}$~\cite{FearickCa48}.

In the oxygen chain, closed-shell CC results align well with experiment in $^{16}\rm{O}$ and $^{22}\rm{O}$.
However, the 2PR ansatz underestimates closed-shell CC predictions in $^{14}\rm{O}$ and $^{22}\rm{O}$, by almost 40\% in the latter case, though the large error bands, dominated by many-body truncation uncertainties, cover much of this discrepancy. 
In $^{20}\rm{O}$, NCSM calculations~\cite{Stumpf2017,StumpfThesis} predict $\alpha_D = 0.66\,(7)\,\rm{fm}^{3}$, consistent with the 2PR ansatz within theoretical uncertainties.  
Between mass numbers 18 and 22, $\alpha_D$ varies relatively little in open-shell CC, contrasting with NCSM predictions, which indicate an increasing trend throughout the isotopic chain, from $\alpha_D = 0.58\,(6)\,\rm{fm}^{3}$ in $^{18}\rm{O}$ to $\alpha_D = 0.79\,(6)\,\rm{fm}^{3}$ in $^{22}\rm{O}$.

For the calcium chain, direct comparisons of 2PR/2PA predictions with experiment or other \textit{ab initio} calculations are unavailable. It might be interesting to compare them to artificial neural network results of Ref.~\cite{Jiang2024AlphaD},
which suggests that both 2PR and 2PA schemes, while consistent with each other,  tend to underestimate $\alpha_D$.

\section{Conclusions and perspectives}
\label{sec: conclusions}

In this paper, we report on progress towards tackling open-shell nuclei within the spherical coupled-cluster formalism. 
In particular, we developed and implemented the spherical two-particle-removed EOM technique. 

First, we validated the 2PR method on g.s.~energies in the oxygen and calcium chains. 
For the $D$$/1p$-$3h$ truncation, we find an accuracy comparable to that of the closed-shell CCSD approach and the $D$$/3p$-$1h$ truncation in the 2PA-EOM framework.
Interestingly, the inclusion of triples in the g.s.~improves the agreement of the different EOM schemes with the experiment considerably, reaching an accuracy similar to that of the closed-shell $T$-1 approximation. 
By combining the particle-attached/removed techniques, a fairly comprehensive description of experimental binding energies across the $\rm{O}$ and $\rm{Ca}$ isotopic chains has been obtained.

Next, we have applied 2PR-EOM to the excited states of $^{14}$C, $^{14}$N and $^{22}$O. 
Our tests suggest that 2PR-EOM at the level of $1p$-$3h$ excitations works well for states with a structure dominated by $2h$ configurations.
For positive-parity states of $^{14}\rm{C}$ and $^{14}\rm{N}$, closed-shell EOM and 2PR-EOM reach comparable accuracies.
Interestingly, the 2PR ansatz performs rather well in $^{22}\rm{O}$, where it is superior to closed-shell CC and correctly describes five excited states when confronted with data.

Finally, we studied electric dipole polarizability by combining the 2PR ansatz with the LIT technique.
We validated the new method by comparing it to the 2PA-LIT ansatz for $^{38}$Ca and found a good agreement within the theoretical uncertainties.
At the same time, a comparison with closed-shell CC predictions confirmed a tendency of both 2PA and 2PR calculations to underestimate $\alpha_D$.
The deviation is significant in the $\rm{Ca}$ chain, and higher-order many-body contributions should be included.

This work testifies to the power of the EOM-CC framework in providing a unified approach for studying the structure and response of open-shell nuclei close to shell subclosures.
Future developments will extend in two directions.
On the one hand, the inclusion of higher-order correlations will be investigated. The 2PR ansatz has been developed in quantum chemistry up to $2p$-$4h$ contributions; its application to nuclear physics requires approximations to keep the computational cost manageable.
On the other hand, we plan to study odd nuclei employing the one-particle-attached/removed schemes.

\section{Acknowledgements}
We thank Weiguang Jiang, Joanna E.~Sobczyk, and Michael Gennari for useful discussions.
This work was supported by the Deutsche Forschungsgemeinschaft (DFG, German Research Foundation) – Project-ID 279384907 – SFB 1245, through the Cluster of Excellence “Precision Physics, Fundamental Interactions, and Structure of Matter” (PRISMA+ EXC 2118/1, Project ID 39083149), and by the U.S. Department of Energy, Office of Science, Office of Nuclear Physics, under the FRIB Theory Alliance award DE-SC0013617, Office of Advanced Scientific Computing Research and Office of Nuclear Physics, and by the U.S. Department of Energy (DOE), Office of Science, under SciDAC-5 (NUCLEI collaboration) and contract DE-FG02-97ER41014. This research used resources of the Oak Ridge Leadership Computing Facility located at Oak Ridge National Laboratory, which is supported by the Office of Science of the Department of Energy under contract No. DE-AC05-00OR22725. Computer time was provided by the Innovative and Novel Computational Impact on Theory and Experiment (INCITE) program and the supercomputer Mogon at Johannes Gutenberg Universit\"at Mainz.

\appendix

\section{Spherical implementation of the 2PR-EOM method}
\label{app: spherical 2pr}
This Appendix provides details on the spherical implementation of the 2PR-EOM-CC method.
We have reported the EOM diagrams in both the standard and angular-momentum-coupled forms in Tab.~\ref{tab: 2pr eom diagrams}.
\begin{table*}[]
    \caption{
    Coupled-cluster diagrams for the right amplitude $\hat{R}^{A-2}$ in the 2PR-EOM method. 
    Both ordinary and reduced expressions are reported. 
    Angular-momentum-coupled amplitudes and Hamiltonian matrix elements are defined in Apps.~\ref{sec: spherical eom amplitudes} and~\ref{sec: spherical hamiltonian}, respectively.
    Wiggly lines denote the similarity-transformed Hamiltonian $\Bar{H}$, while the $0p$-$2h$ and $1p$-$3h$ $R$ amplitudes are shown as bold horizontal lines.
    Explicit three-body terms in $\Bar{H}$ are neglected.
    }
    \begin{tabular}
    {>{\raggedright} m{4cm} m{4cm} m{8cm} }
        Diagram & Uncoupled expression & Coupled expression \\
        \hline
        \noalign{\vskip 2mm}   
        % hh
        \centering\raisebox{-0.5\height}
        { \includegraphics[width=0.25\columnwidth]{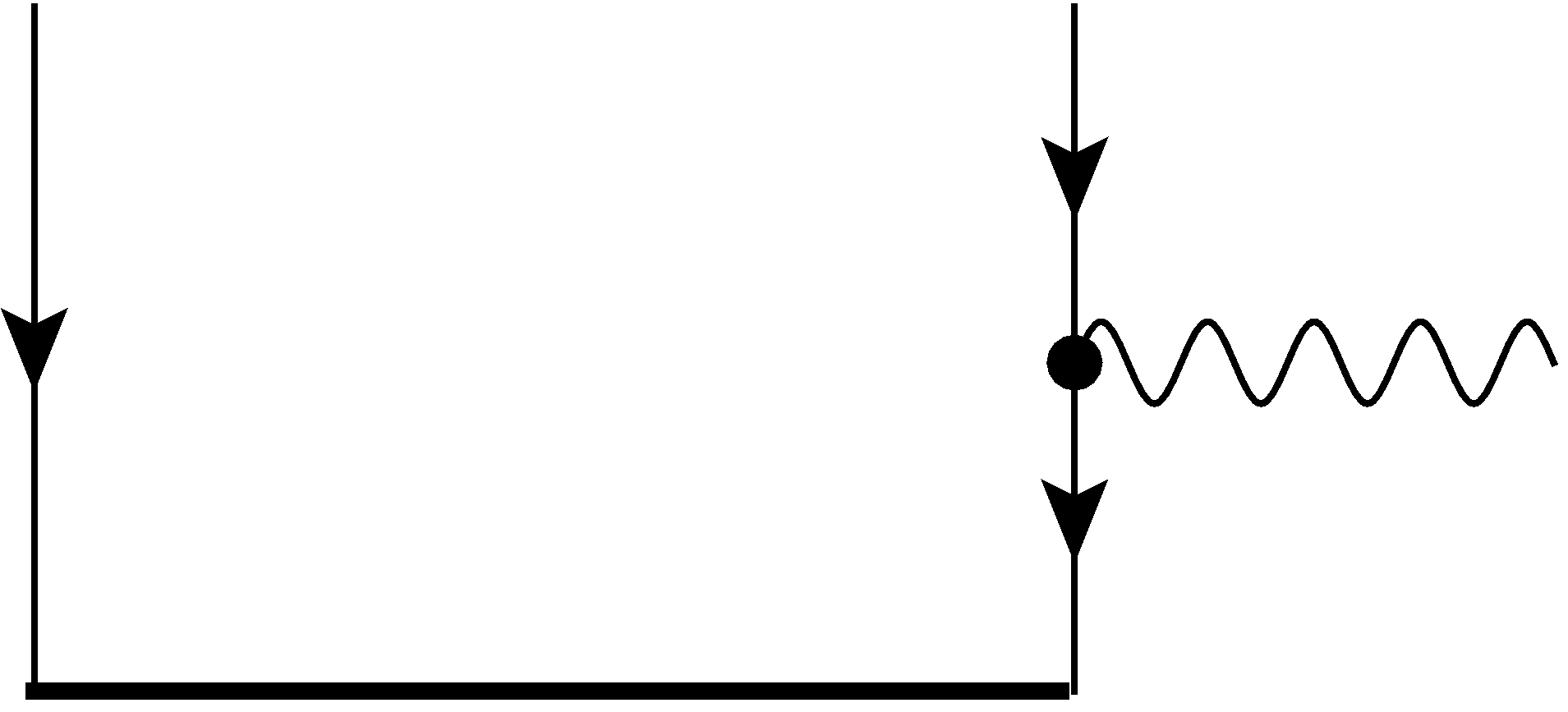} } & 
        $-P(ij) \sum_m r_{im} \Bar{H}^{m}_{j}$ & 
        $-P(ij) \sum_m \delta_{j_j j_m} r_{im}(J) \Bar{H}^{m}_{j}(j_j)$ 
        \\
        \vskip 4mm
        % hhhh
        \centering\raisebox{-0.5\height}
        { \includegraphics[width=0.25\columnwidth]{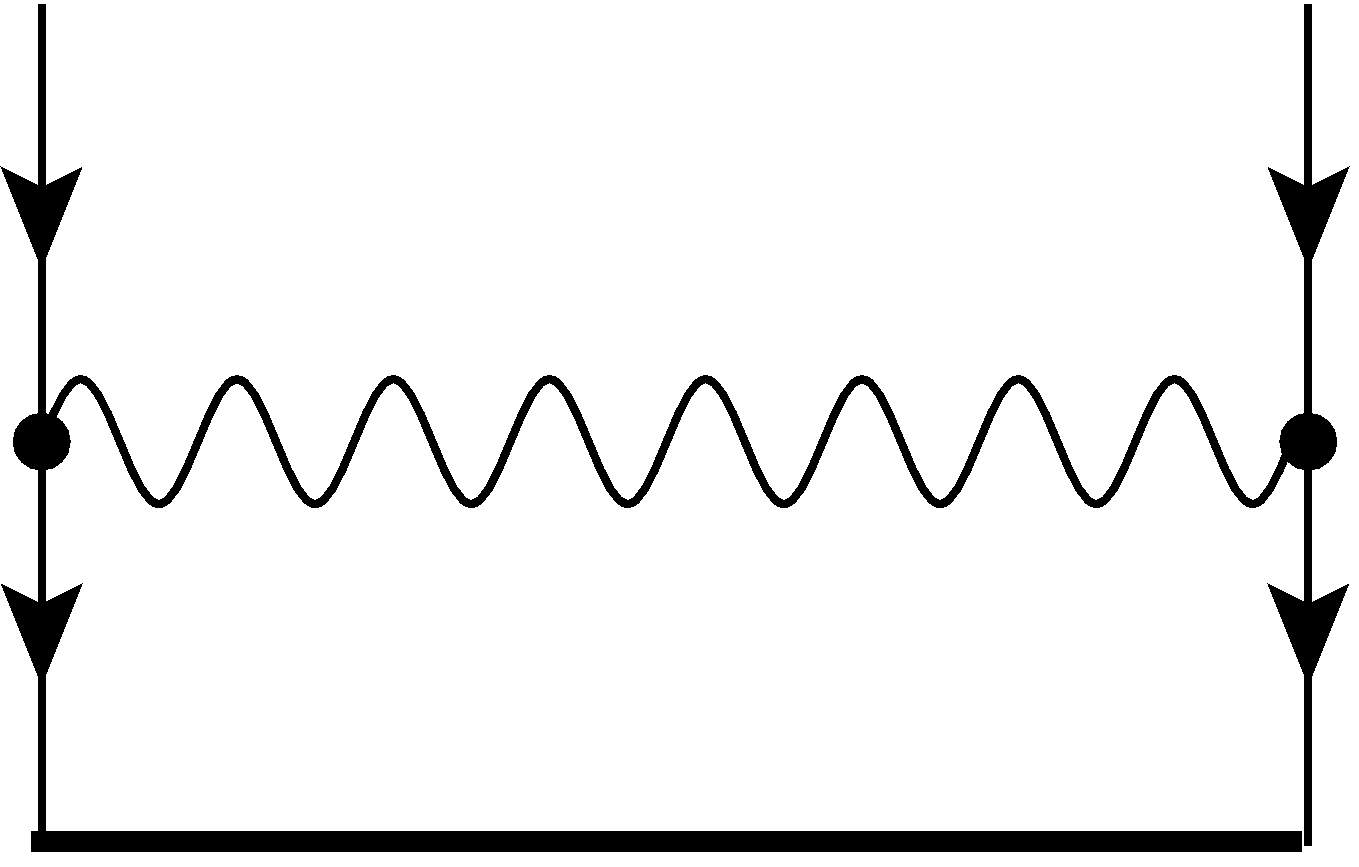} } &
        $ \frac{1}{2} \sum_{mn} r_{mn} \Bar{H}^{mn}_{ij} $ &
        $ \frac{1}{2} \sum_{mn} r_{mn}(J) \Bar{H}^{mn}_{ij}(J) $ 
        \\
        \vskip 4mm
        % ph-hh
        \centering\raisebox{-0.5\height}
        { \includegraphics[width=0.25\columnwidth]{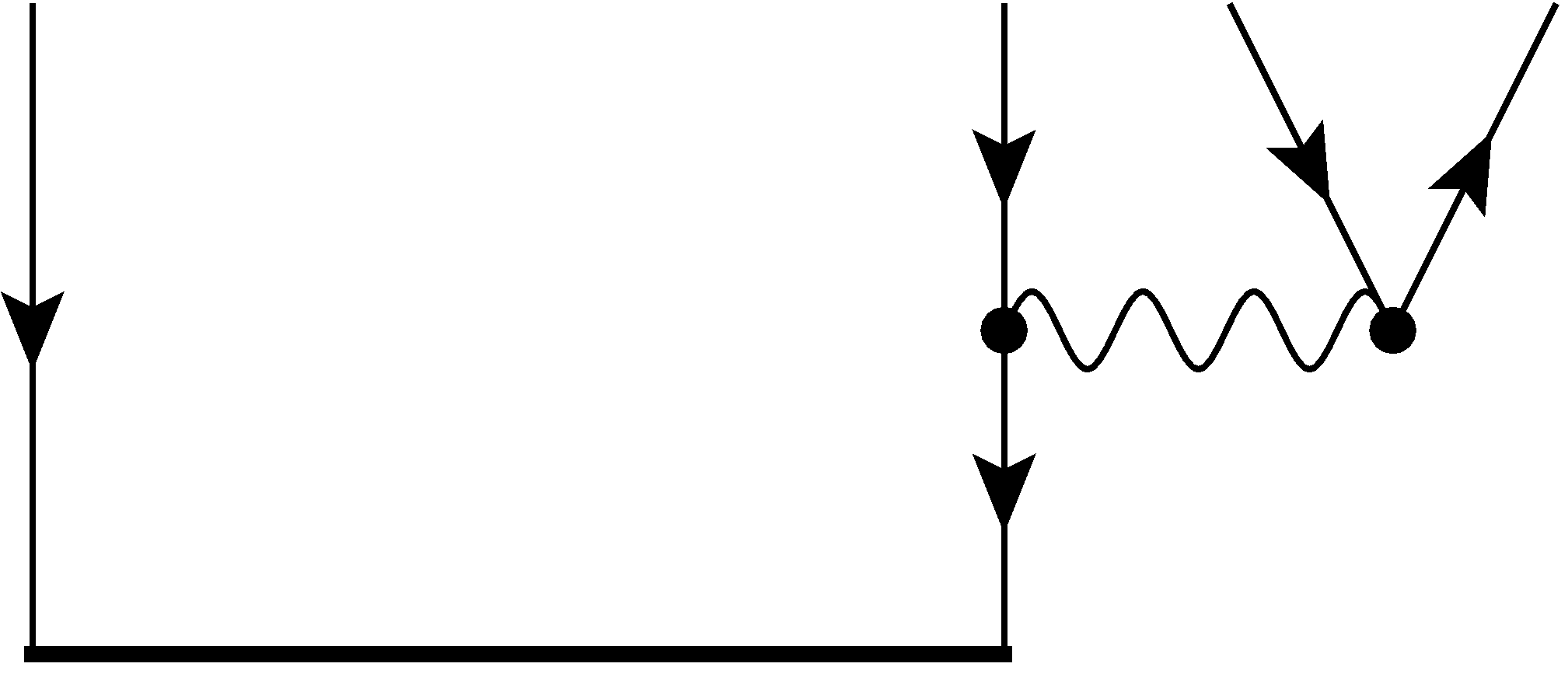} } &
        $ - P(ij,k) \sum_m r_{km} \Bar{H}^{am}_{ij} $ &
        $ - P(ij,k) \, \frac{ \hat{J}_{ij} \hat{J}_{ijk} }{ \hat{j}_a } \, \sum_m \, (-1)^{ j_a+j_k +j_m + J_{ijk} } $ \\
        & & 
        $ \qquad \times \, 
        \begin{Bmatrix}
            j_a & j_m & J_{ij} \\ j_k & J_{ijk} & J
        \end{Bmatrix}
        \, 
        r_{km}(J) \Bar{H}^{am}_{ij}(J_{ij})
        $
        \\
        \vskip 4mm
        % ph
        \centering\raisebox{-0.5\height}
        { \includegraphics[width=0.25\columnwidth]{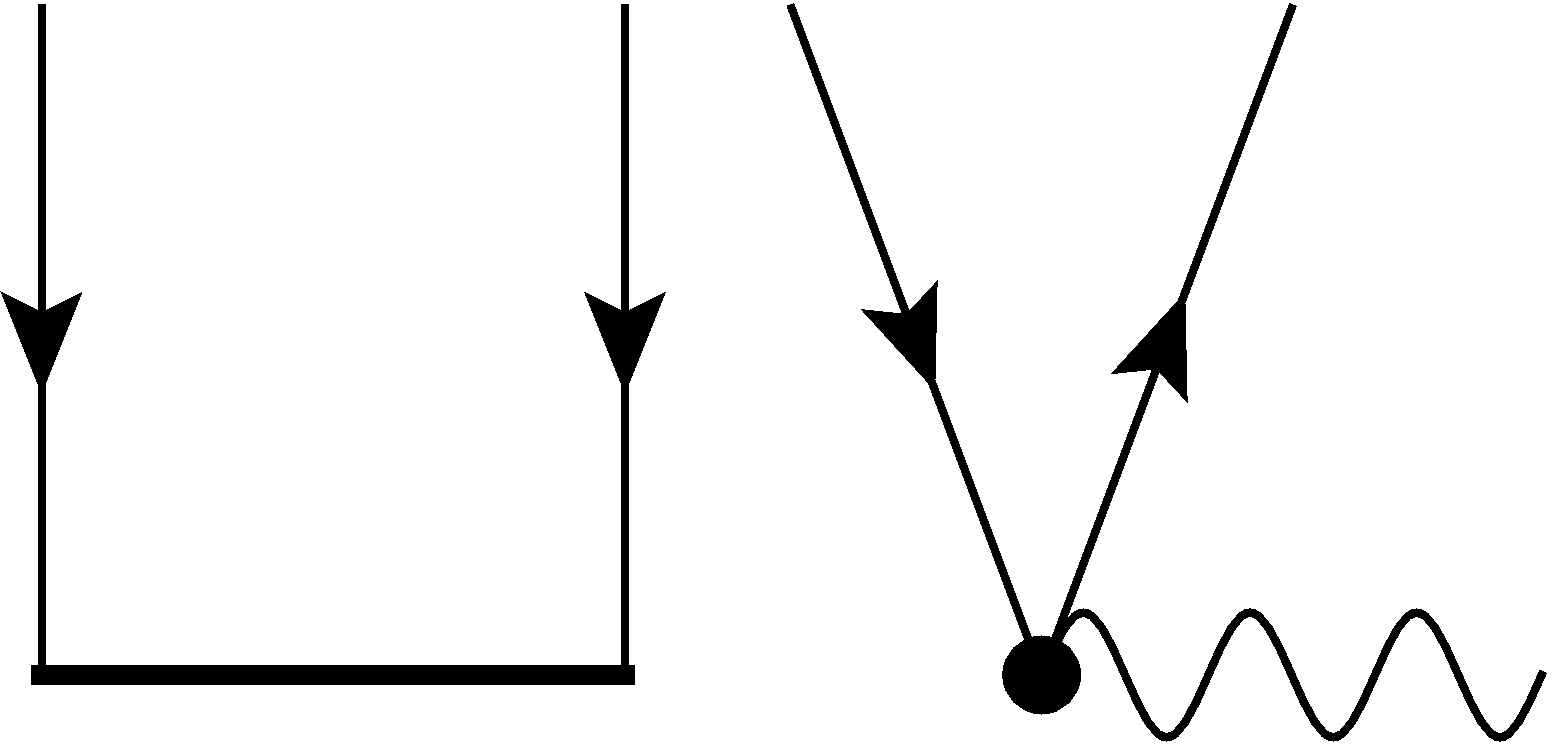} } &
        $ P(ij,k) \, r_{ij} \Bar{H}^{a}_{k} $ &
        $ P(ij,k) \,
        (-1)^{ j_a + J - J_{ijk} } \,
        \frac{ \hat{J}_{ijk} }{ \hat{j}_a \hat{J} } \,
        r_{ij}(J) \Bar{H}^{a}_{k}(j_a) $
        \\
        \vskip 4mm
        % hp
        \centering\raisebox{-0.5\height}
        { \includegraphics[width=0.25\columnwidth]{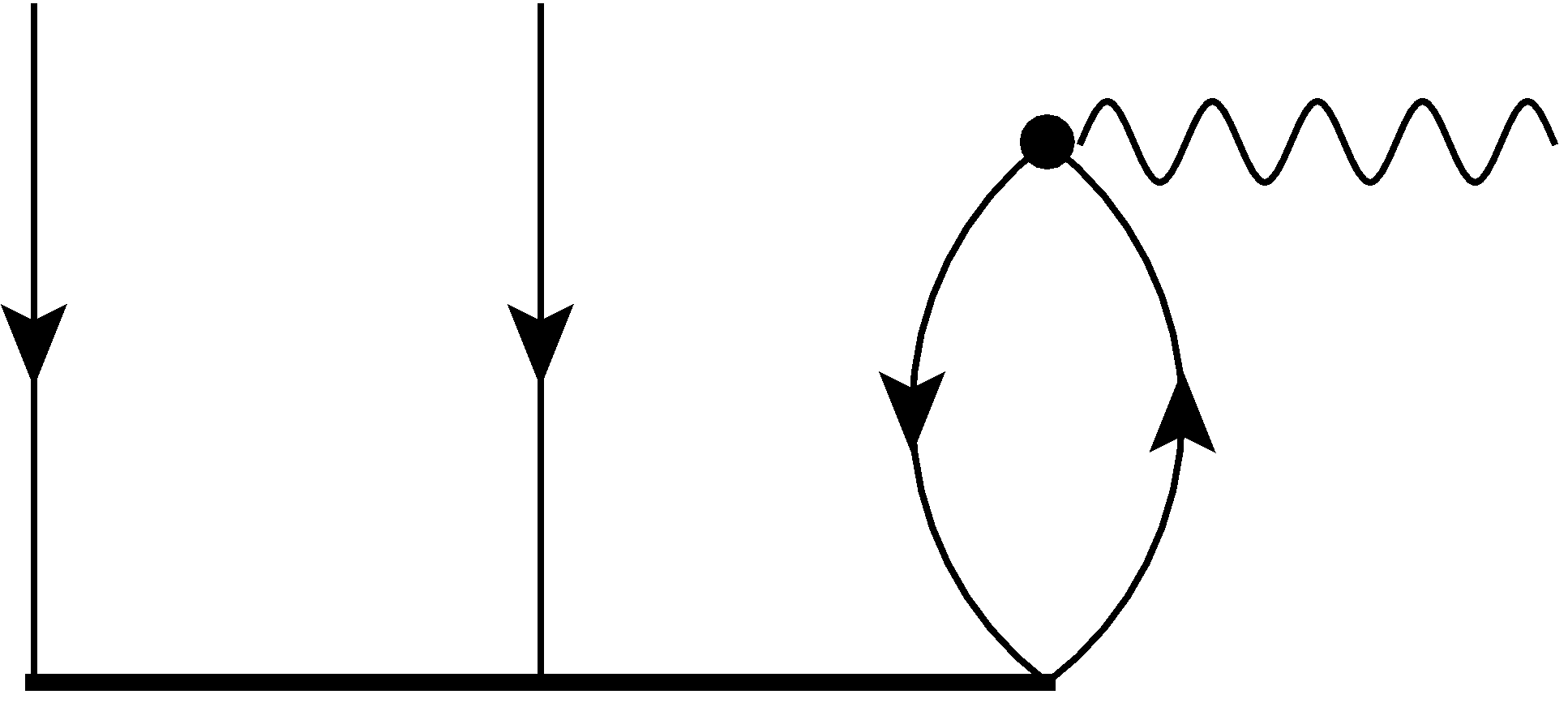} } &
        $\sum_{em} r^{e}_{ijm} \Bar{H}^{m}_{e}$ &
        $\sum_{em, J_{ijm} } \,
        (-1)^{ J_{ijm} - j_m - J} \,
        \frac{ \hat{j}_m \hat{J}_{ijm} }{ \hat{J} }
        $ 
        \\ 
        \vskip 4mm 
        & &  $ \qquad \times \, r^{e}_{ijm}(J, J_{ij},J_{ijm}) \Bar{H}^{m}_{e}(j_m)$ 
        \\
        \vskip 4mm
        % hh-hp
        \centering\raisebox{-0.5\height}
        { \includegraphics[width=0.25\columnwidth]{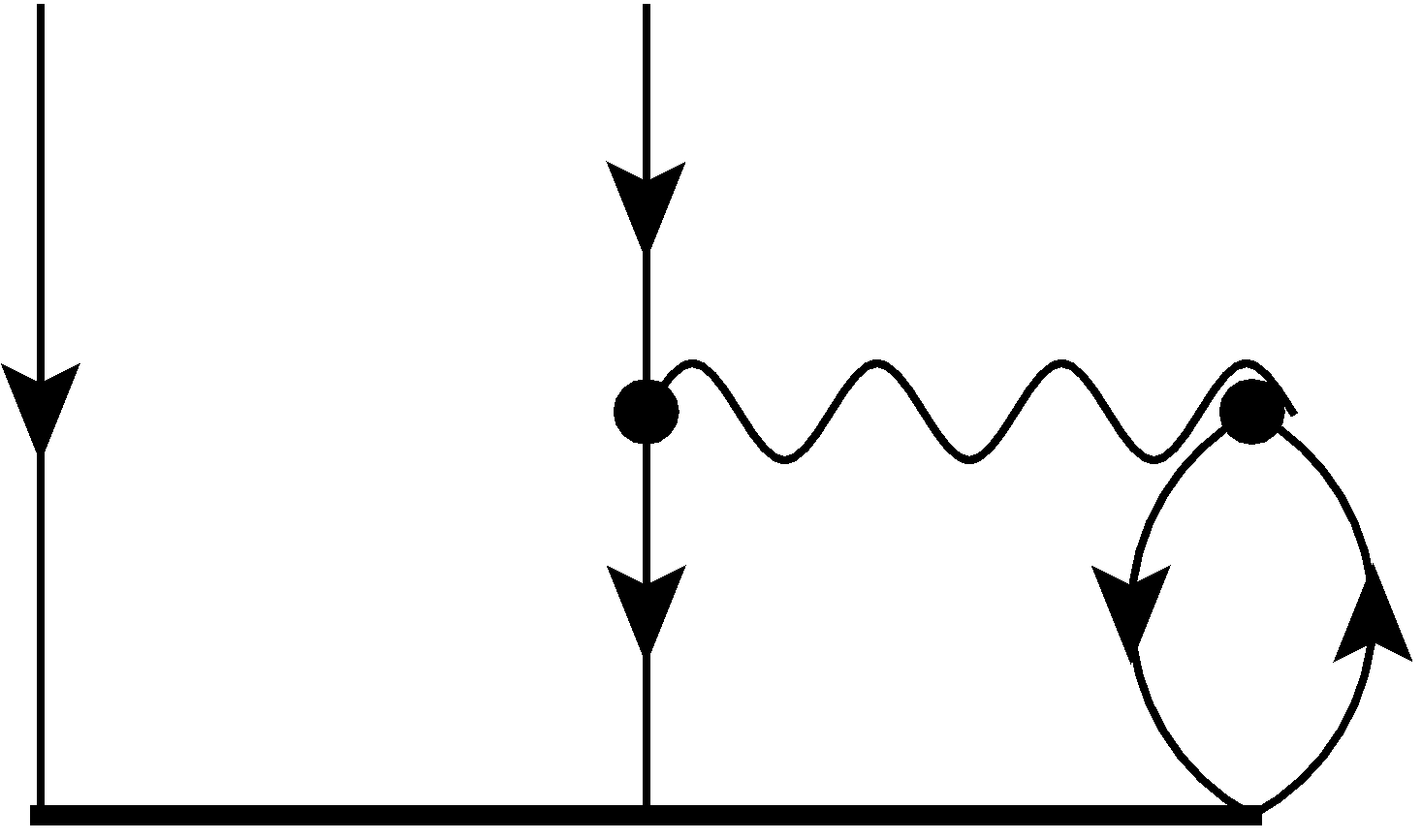} } &
        $ \frac{1}{2} P(ij) \sum_{emn} r^{e}_{mnj} \Bar{H}^{mn}_{ie}$ &
        $ \frac{1}{2} P(ij) \sum_{emn, J_{mn} J_{mnj} } \,
        (-1)^{ j_j + J_{mn} + J_{mnj} } \,
        \hat{j}_e \hat{J}_{mn} \hat{J}_{mnj} $ \\
        & & $
        \qquad \times \,
        \begin{Bmatrix}
            j_e & j_i & J_{mn} \\ j_j & J_{mnj} & J
        \end{Bmatrix}
        r^{e}_{mnj}(J,J_{mn}, J_{mnj}) \Bar{H}^{mn}_{ie}(J_{mn})$
        \\
        \vskip 4mm
        % hh
        \centering\raisebox{-0.5\height}
        { \includegraphics[width=0.25\columnwidth]{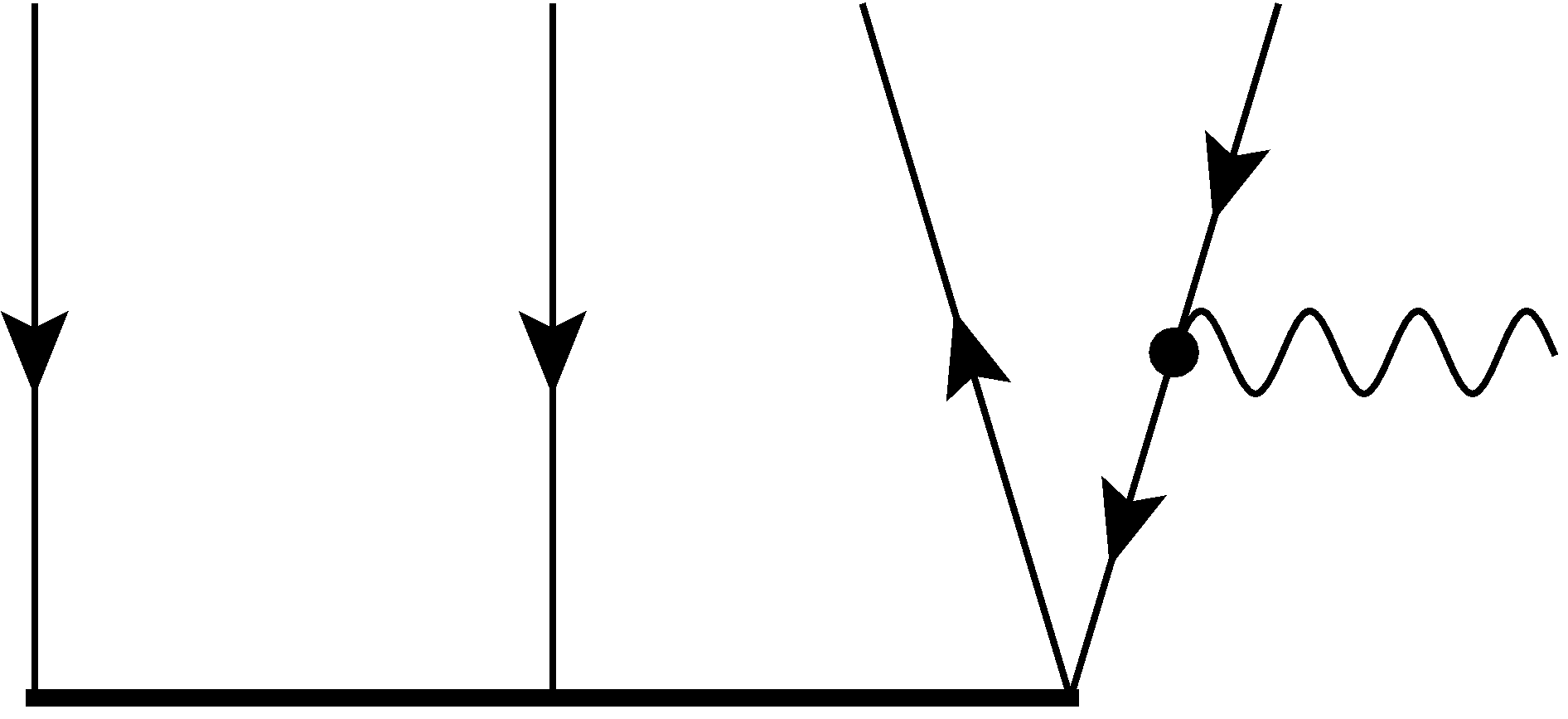} } &
         $ -P(ij,k) \sum_{m} r^{a}_{ijm} \Bar{H}^{m}_{k} $ &
         $ -P(ij,k) \sum_{m} r^{a}_{ijm}(J, J_{ij}, J_{ijk}) \Bar{H}^{m}_{k}(j_k) $ 
        \\
        \vskip 4mm
        % pp
        \centering\raisebox{-0.5\height}
        { \includegraphics[width=0.25\columnwidth]{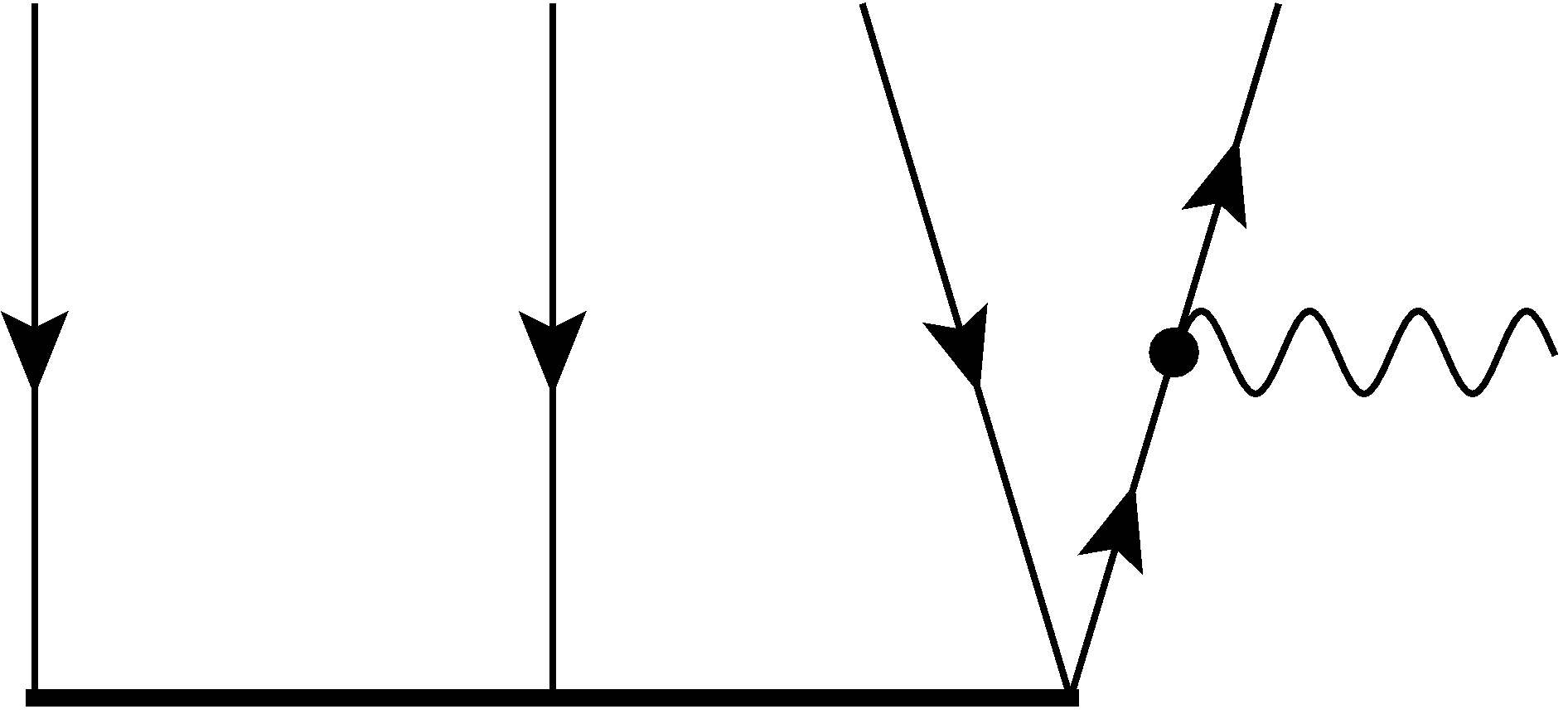} } &
         $\sum_e r^{e}_{ijk} \Bar{H}^{a}_{e} $ &
         $\sum_e r^{e}_{ijk}(J, J_{ij}, J_{ijk})  \Bar{H}^{a}_{e}(j_a) $ 
         \\
         \vskip 4mm
        % hhhh
        \centering\raisebox{-0.5\height}
        { \includegraphics[width=0.25\columnwidth]{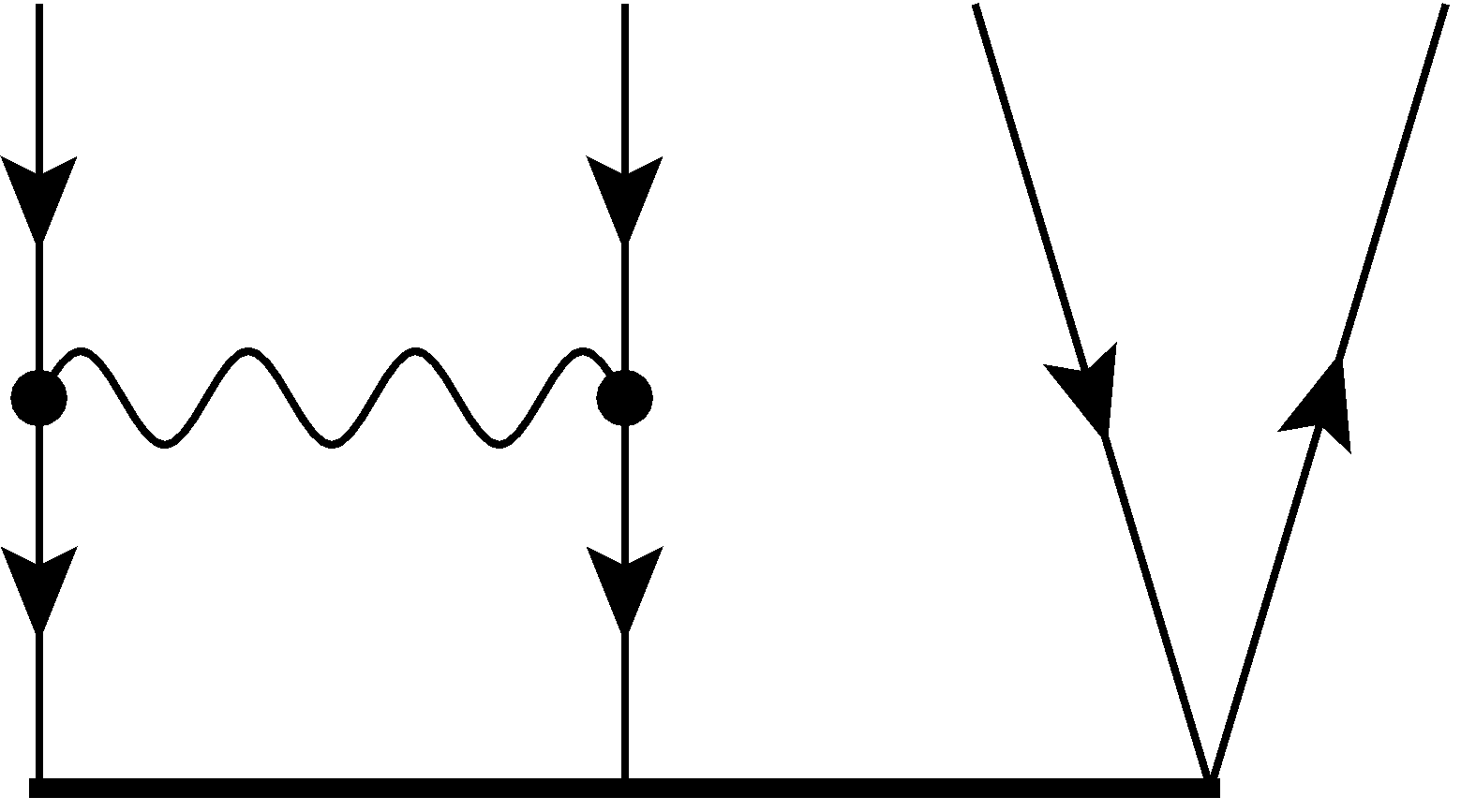} } &
         $ \frac{1}{2} P(ij,k) \sum_{mn} r^{a}_{mnk} \Bar{H}^{mn}_{ij} $ &
         $ \frac{1}{2} P(ij,k) \sum_{mn} r^{a}_{mnk}(J, J_{ij}, J_{ijk}) \Bar{H}^{mn}_{ij}( J_{ij} ) $ 
         \\
         \vskip 4mm
        % hp-hp
        \centering\raisebox{-0.5\height}
        { \includegraphics[width=0.25\columnwidth]{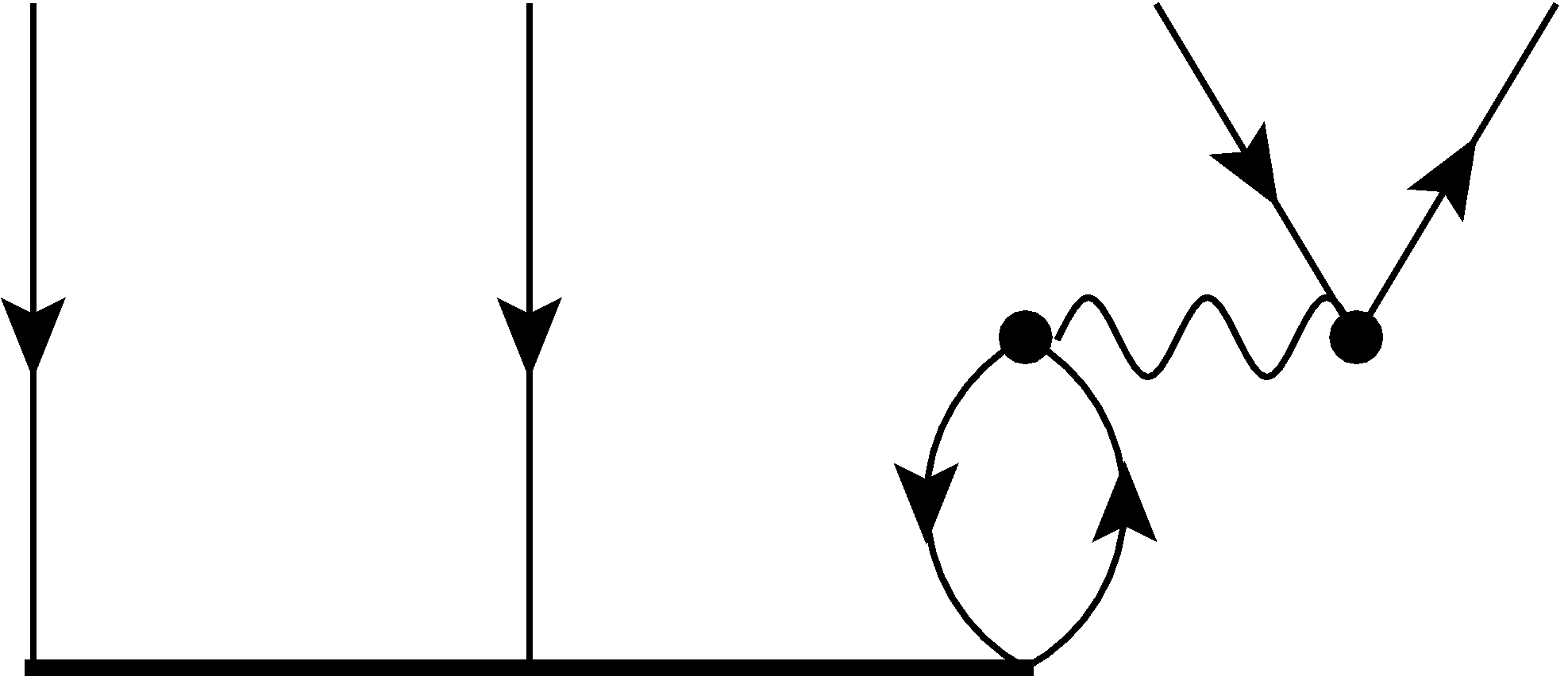} } &
         $ - P(ij,k) \sum_{em} r^{e}_{ijm} \Bar{H}^{ma}_{ke}  $ &
         $ - P(ij,k) \sum_{em, J_{ma} J_{ijm} } 
         (-1)^{ j_a + j_e + j_k + j_m } \,
         \frac{ \hat{j}_e \hat{J}_{ma}^2 \hat{J}_{ijk} \hat{J}_{ijm} }{ \hat{j}_{a} }
         $
         \\ & & $
         \qquad \times \,
         \begin{Bmatrix}
            J & J_{ijm} & j_{e} \\ J_{ijk} & J_{ij} & j_k \\ j_a & j_m & J_{ma}
        \end{Bmatrix}
         r^{e}_{ijm}( J, J_{ij}, J_{ijk}) \Bar{H}^{ma}_{ke}(J_{ma}) $ 
    \end{tabular}
    \label{tab: 2pr eom diagrams}
\end{table*}

We introduce our notation for the reduced matrix elements of spherical tensor operators.
The Wigner-Eckart theorem allows one to factor out the dependence on the projection quantum numbers in the matrix elements of a spherical tensor operator, with total angular momentum $J$ and projection $M$, between states of well-defined angular momentum.
We follow the convention of Refs.~\cite{Jansen2013,Bonaiti2024,BonaitiThesis} and define the reduced matrix elements for the right amplitudes and the Hamiltonian as follows
\begin{align}
    \label{eq: wet right}
    \mel{0}{\hat{R}^{JM}}{ij} = C^{00}_{JM, J_{ij} M_{ij} }
    \big\langle 0 || \hat{R}^{J} || ij \big\rangle,
\end{align}
and, for the left amplitudes, as
\begin{align}
    \label{eq: wet left}
    \mel{ij}{\hat{L}^{JM}}{0} = C^{00}_{JM, J_{ij} M_{ij} }
    \big\langle ij || \hat{L}^{J} || 0 \big\rangle.
\end{align}

Transformations between the $m$-scheme and $J$-scheme tensors are discussed below, see also Ref.~\cite{Jansen2013}.
In the $m$-scheme expressions, the single-particle indices denote the set of quantum numbers 
\begin{align}
    \ket{p} = \ket{ n_p (l_p \frac{1}{2})  j_p m_p; \frac{1}{2} t_{z,p} } = \ket{ \alpha_p; j_p m_p},
\end{align}
with orbital angular momentum $l_p$ and spin coupled to $j_p$, isospin projection $t_{z,p}$, and principal quantum number $n_p$.
In $J$-scheme, instead, the dependence on the projection $m_p$ is dropped, and the label $p$ must be understood as $\ket{p} = \ket{\alpha_p; j_p}$.

\subsection{Spherical matrix elements of the Hamiltonian}
\label{sec: spherical hamiltonian}

We use the following compact notation for the matrix elements of the similarity-transformed Hamiltonian $\Bar{H}$:
\begin{align}
    \Bar{H}^{p}_{q} &= \mel{p}{ \Bar{H}}{q}, \\
    \Bar{H}^{pq}_{rs} &= \mel{pq}{ \Bar{H} }{rs}_{A},
\end{align}
with the subscript $A$ denoting that matrix elements are antisymmetrized.
Indices  $pqr\, ... $ refer to generic s.p. states. 

$\Bar{H}$ is a scalar operator.
Its one-body components in the coupled and uncoupled representations are given by
\begin{align}
    \label{eq: H 1b dir}
    & \Bar{H}^{p}_{q} = \delta_{ j_p j_q } \delta_{ m_p m_q }  \Bar{H}^{p}_{q}(j_p), 
    \\ 
    \label{eq: H 1b inv}
    & \Bar{H}^{p}_{q}(j_p) = \delta_{ j_p j_q } \Bar{H}^{p}_{q}.
\end{align}
For the two-body matrix elements,
\begin{align}
    \label{eq: H 2b dir}
    & \Bar{H}^{pq}_{rs} = \sum_{J_{qp} M_{pq} } 
    C^{ J_{qp} M_{pq} }_{ j_p m_p, j_q m_q } 
    C^{ J_{qp} M_{pq} }_{ j_r m_r, j_s m_s }
    \Bar{H}^{pq}_{rs}( J_{pq} ), \\
    \label{eq: H 2b inv}
    & \Bar{H}^{pq}_{rs}( J_{pq} ) = \frac{1}{ \hat{J}_{pq}^{2} } \,
    \sum_{ \substack{ m_p m_q \\ m_r m_s M_{pq} } }
    C^{ J_{qp} M_{pq} }_{ j_p m_p, j_q m_q } 
    C^{ J_{qp} M_{pq} }_{ j_r m_r, j_s m_s }
    \Bar{H}^{pq}_{rs}
\end{align}
having defined the reduced matrix elements 
\begin{align}
    \Bar{H}^{p}_{q}(j_p) &=
    \big\langle {p; j_p}
    || \Bar{H} ||
    {q; j_q = j_p}
     \big\rangle
    \\
    \Bar{H}^{pq}_{rs}( J_{pq} ) &=
    \big\langle
    {pq; j_p j_q; J_{pq} }
    || \Bar{H} ||
    { rs; j_{r} j_{s}; J_{rs}=J_{pq} }
    \big\rangle.
\end{align}

\subsection{Spherical 2PR-EOM amplitudes}
\label{sec: spherical eom amplitudes}

The 2PR-EOM amplitudes are generally spherical tensors carrying quantum numbers $J$, $M$.
The right 2PR excitation operator $\hat{R}$, defined in Eq.~\eqref{eq: eom 2pr right}, features $0p$-$2h$ and $1p$-$3h$ contributions,
\begin{align}
    r_{ij} &= \mel{ \Phi_{ij} }{ \hat{R}^{JM} }{ \Phi_0} =
    \mel{ 0 }{ \hat{R}^{JM} }{ ij }, \\
    r_{ijk}^{a} &= \mel{ \Phi_{ijk}^{a} }{ \hat{R}^{JM} }{ \Phi_0} =
    \mel{ a }{ \hat{R}^{JM} }{ ijk },
\end{align}
respectively. Similar expressions hold for the left amplitudes $l^{ij}$ and $l^{ijk}_{a}$.

Starting from Eqs.~\eqref{eq: wet right} and~\eqref{eq: wet left}, respectively, we find for the right and left amplitudes $\hat{R}^{JM}$ and $\hat{L}^{JM}$
\begin{align}
    & \mel{0}{ \hat{R}^{JM} }{ ij }  
    \\ 
    & \qquad = \sum_{J_{ij} M_{ij} }
    C_{ j_i m_i, j_j m_j}^{ J_{ij} M_{ij} } C_{JM, J_{ij} M_{ij} }^{00} 
    \left\langle 
    0
    \left\| R^{J} \right\| 
    ij; j_i j_j; J_{ij} 
    \right\rangle 
    \nonumber
    \\
    & \qquad = \frac{ (-1)^{J-M} }{ \hat{J} } C_{ j_i m_i, j_j m_j}^{ J, -M }
    r_{ij}(J) ,
    \nonumber
    \\ 
    & \mel{ ij }{ \hat{L}^{JM} }{ 0 }  =
    \frac{ (-1)^{J-M} }{ \hat{J} } C_{ j_i m_i, j_j m_j}^{ J, -M }
    l^{ij}(J),
\end{align}
where the shorthand notation 
\begin{align}
    r_{ij}(J) &= \left\langle 
    0
    \left\| R^{J} \right\| 
    ij; j_i j_j; J_{ij} = J
    \right\rangle,
    \\
    l^{ij}(J) &= \left\langle 
    ij; j_i j_j; J_{ij} = J
    \left\| L^{J} \right\| 
    0
    \right\rangle 
\end{align}
has been introduced, $\hat{J} = \sqrt{2J+1}$, and $j_i$ and $j_j$ have been coupled to total angular momentum $J_{ij}$.
Relation to express the $0p$-$2h$ reduced elements as a function of the $m$-scheme elements read
\begin{align}
    \label{eq: R 0p2h inv}
    & r_{ij}(J) = \frac{1}{\hat{J}} \sum_{m_i m_j M} (-1)^{J-M} C^{J, -M}_{ j_i m_i, j_j m_j } r_{ij},    \\
    \label{eq: L 0p2h inv}
    & l^{ij}(J) = \frac{1}{\hat{J}} \sum_{m_i m_j M} (-1)^{J-M} C^{J, -M}_{ j_i m_i, j_j m_j } l^{ij},
\end{align}

When three angular momenta appear either on the bra or ket side, e.g. $\ket{ijk}$, by convention we always couple them from the left to the right, namely, first we couple $j_i$ to $j_j$ to $J_{ij}$, then $J_{ij}$ to $j_k$ to form $J_{ijk}$.
Then, transformations for the $1p$-$3h$ amplitudes are listed below:
\begin{align}
    \label{eq: R 1p3d dir}
    & r^{a}_{ijk} =
    \sum_{ \substack{ J J_{ij} J_{ijk} \\ M M_{ij} M_{ijk} } } 
    C^{ J_{ij} M_{ij} }_{ j_i m_i, j_j m_j } 
    C^{ J_{ijk} M_{ijk} }_{ J_{ij} M_{ij}, j_k m_k }
     \nonumber
    \\ 
    & \qquad  \qquad \times \, 
    C^{ j_{a} m_{a} }_{JM, J_{ijk} M_{ijk} }
    r^{a}_{ijk}(J, J_{ij}, J_{ijk} ),
    \\
    \label{eq: R 1p3d inv}
    & r^{a}_{ijk}(J, J_{ij}, J_{ijk} )
    = \frac{1}{ \hat{j}_{a}^2 } \sum_{ \substack{ m_a m_i m_j m_k \\ M M_{ij} M_{ijk} } }
    C^{ J_{ij} M_{ij} }_{ j_i m_i, j_j m_j } 
    C^{ J_{ijk} M_{ijk} }_{ J_{ij} M_{ij}, j_k m_k } 
    \nonumber
    \\ & \qquad  \qquad \times \,
    C^{ j_{a} m_{a} }_{JM, J_{ijk} M_{ijk}} \,
    r^{a}_{ijk},
\end{align}
where
\begin{align}
    & r^{a}_{ijk}(J, J_{ij}, J_{ijk} ) = \\ 
    & \qquad \left\langle 
    a; j_a
    \left\| R^{J} \right\| 
    ijk; j_i j_j; J_{ij} j_k ; J_{ijk}
    \right\rangle.
    \nonumber
\end{align}

Left $1p$-$3h$ amplitudes are given by
\begin{align}
    \label{eq: L 1p3h dir}
    & l_{a}^{ijk} =
    \sum_{ \substack{ J J_{ij} J_{ijk} \\ M M_{ij} M_{ijk} } } 
    C^{ J_{ij} M_{ij} }_{ j_i m_i, j_j m_j } 
    C^{ J_{ijk} M_{ijk} }_{ J_{ij} M_{ij}, j_k m_k }
     \nonumber
    \\ 
    & \qquad  \qquad \times \, 
    C^{ j_{a} m_{a} }_{JM, J_{ijk} M_{ijk} }
    l_{a}^{ijk}(J, J_{ij}, J_{ijk} ),
    \\
    \label{eq: L 1p3h inv}
    & l_{a}^{ijk}(J, J_{ij}, J_{ijk} )
    = \frac{1}{ \hat{j}_{a}^2 } \sum_{ \substack{ m_a m_i m_j m_k \\ M M_{ij} M_{ijk} } }
    C^{ J_{ij} M_{ij} }_{ j_i m_i, j_j m_j } 
    C^{ J_{ijk} M_{ijk} }_{ J_{ij} M_{ij}, j_k m_k } 
    \nonumber
    \\ & \qquad  \qquad \times \,
    C^{ j_{a} m_{a} }_{JM, J_{ijk} M_{ijk}} \,
    l_{a}^{ijk},
\end{align}
where
\begin{align}
    & l_{a}^{ijk}(J, J_{ij}, J_{ijk} ) = \\ 
    & \qquad \left\langle 
    ijk; j_i j_j; J_{ij} j_k ; J_{ijk}
    \left\| L^{J} \right\| 
    a; j_a
    \right\rangle.
    \nonumber
\end{align}

\section{Spherical 2PR-LIT}
\label{sec: spherical 2pr lit}
The spherical formulation of the 2PR-LIT-CC method is detailed in this section.
The similarity-transformed excitation operators are tensors of angular momentum $K$ and projection $M_K$.
$\bar{\Theta}^{KM_K}$ comprises one- and two-body contributions, denoted as $\bar{\Theta}^{p}_{q}$ and $\bar{\Theta}^{pq}_{rs}$, which have been derived in Ref.~\cite{Miorellithesis}.
The corresponding reduced matrix elements are defined as
\begin{align}
    \bar{\Theta}^{p}_{q}(K) &= 
    \left\langle 
    p; j_p
    \left\| \bar{\Theta}^{K} \right\| 
    q; j_q
    \right\rangle, &
    \\ 
    \bar{\Theta}^{pq}_{rs}(K, J_{pq}, J_{rs} ) &= 
    \left\langle 
    pq; j_p j_q; J_{pq}
    \left\| \bar{\Theta}^{K} \right\| 
    rs; j_r j_s; J_{rs}
    \right\rangle,
\end{align}
and are related to the uncoupled matrix elements by the following expressions for the one-body terms,
\begin{align}
    \label{eq: theta 1b dir}
    & \bar{\Theta}^{p}_{q} = 
    C_{KM_K, j_q m_q}^{j_p m_p} \bar{\Theta}^{p}_{q}(K), \\
    & \bar{\Theta}^{p}_{q}(K) = 
    \frac{1}{ \hat{j}_{p}^{2} } \sum_{m_p m_q M_K}
    C_{KM_K, j_q m_q}^{j_p m_p} \bar{\Theta}^{p}_{q}
\end{align}
and two-body terms,
\begin{align}
    & \bar{\Theta}^{pq}_{rs} =
    \sum_{ \substack{J_{pq} M_{pq} \\ J_{rs} M_{rs} } }
    C^{ J_{pq} M_{pq} }_{j_p m_p, j_q m_q} 
    C^{ J_{rs} M_{rs} }_{j_r m_r, j_s m_s} 
    \nonumber
    \\
    & \qquad \times \,
    C^{ J_{pq} M_{pq} }_{ KM, J_{rs} M_{rs}}
    \bar{\Theta}^{pq}_{rs}(K, J_{pq}, J_{rs} ), \\
    & \bar{\Theta}^{pq}_{rs}(K, J_{pq}, J_{rs} ) =
    \frac{1}{ \hat{J}_{pq}^{2} }
    \sum_{\substack{  m_p m_q m_r m_s \\ M_{pq} M_{rs} M_K } }
    C^{ J_{pq} M_{pq} }_{j_p m_p, j_q m_q} 
    C^{ J_{rs} M_{rs} }_{j_r m_r, j_s m_s} 
    \nonumber
    \\
    \label{eq: theta 2b inv}
    & \qquad \times \,
    C^{ J_{pq} M_{pq} }_{ KM, J_{rs} M_{rs}}
    \bar{\Theta}^{pq}_{rs} .
\end{align}

The Lanczos pivots $\mathbf{S}^{R} = \Bar{\Theta} \hat{R}_{0} \ket{\Phi_0}$ and $\mathbf{S}^{L} = \bra{\Phi_0}\hat{L}_{0} \overline{ \Theta^{\dagger} }$ have been derived under the assumption that the g.s.~ of the 2PR nucleus is spherical. 
This hypothesis is valid for a large number of 2PR nuclei and translates into the g.s.~EOM amplitudes, denoted as $\hat{R}_{0}$ and $\hat{L}_{0}$, being scalar tensors ($J=0$).
Therefore, the tensor products $\Bar{\Theta}^K \otimes \hat{R}_0^{(A-2)}$ and $ \hat{L}_0^{(A-2)} \otimes  \overline{ \Theta^{\dagger} }^{K} $ entering the Lanczos pivots become spherical tensors of rank $K$ and thus satisfy the relations Eqs.~\eqref{eq: theta 1b dir}-\eqref{eq: theta 2b inv}.
The diagrams for the right and left pivots are shown in Tab.~\ref{tab: 2pr LIT diagrams} and Tab.~\ref{tab: 2pr LIT diagrams Left}, respectively.

\begin{table*}[h]
    \caption{
    Coupled-cluster diagrams for the right pivot $\mathbf{S}^{R}$ in the 2PR-LIT approach.
    Both ordinary and reduced expressions are reported. 
    Curly lines denote the effective transition operator $\Bar{\Theta}$; double horizontal lines refer to the g.s.~ amplitudes $\hat{R}_{0}$ of the 2PR nucleus.
    }
    \begin{tabular}
    {>{\raggedright} m{4cm} m{4cm} m{8cm} }
        Diagram & Uncoupled expression & Coupled expression \\
        \hline
        \noalign{\vskip 2mm}   
        % hh
        \centering\raisebox{-0.5\height}
        { \includegraphics[width=0.25\columnwidth]{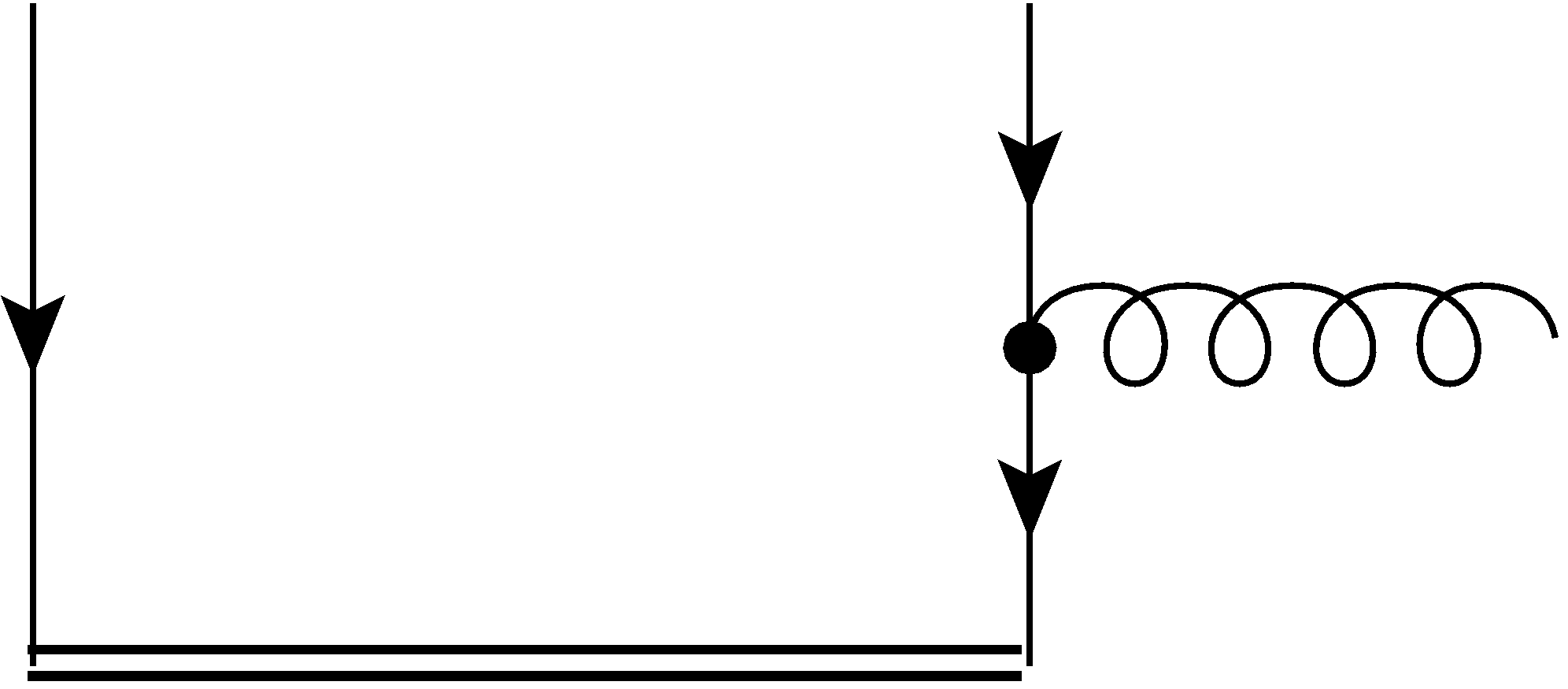} } & 
        $-P(ij) \sum_m (r_0)_{im} \Bar{\Theta}^{m}_{j}$ & 
        $-P(ij) (-1)^{j_i - j_j - K} \sum_m (r_0)_{im}(J_{im}=0) \Bar{\Theta}^{m}_{j}(K) $ 
        \\
        \vskip 4mm
        % hp
        \centering\raisebox{-0.5\height}
        { \includegraphics[width=0.25\columnwidth]{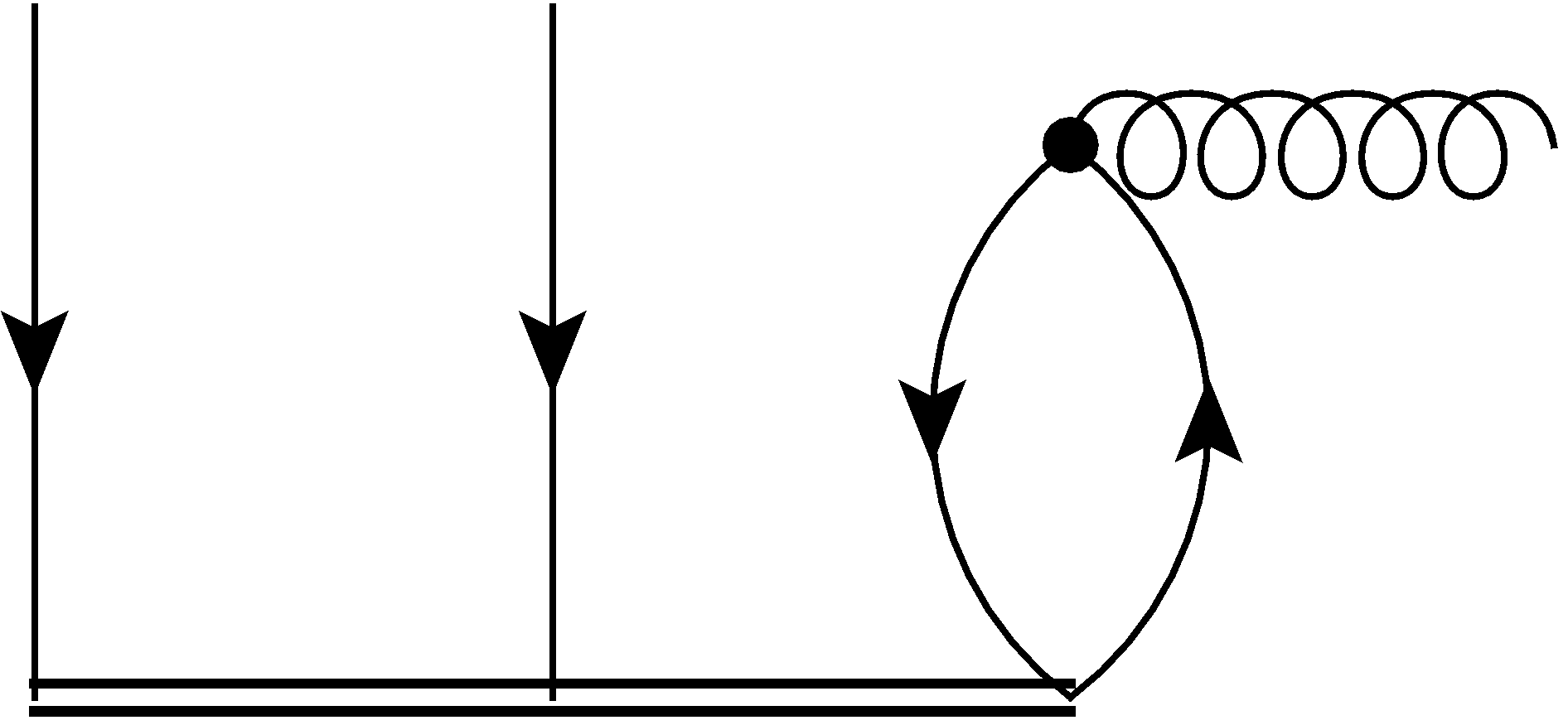} } 
        & $ P(ij) \sum_{dm} (r_0)^{d}_{ijm} \Bar{\Theta}^{m}_{d}$
        & $ P(ij) \sum_{dm}
        \frac{ \hat{j}_d \hat{j}_m }{ \hat{K} } \,
        (-1)^{j_m + K - j_d} \,
        (r_0)^{d}_{ijm}(J_{ij}=K) \, \Bar{\Theta}^{m}_{d}(K) $
        \\
        \vskip 4mm
        % ph
        \centering\raisebox{-0.5\height}
        { \includegraphics[width=0.25\columnwidth]{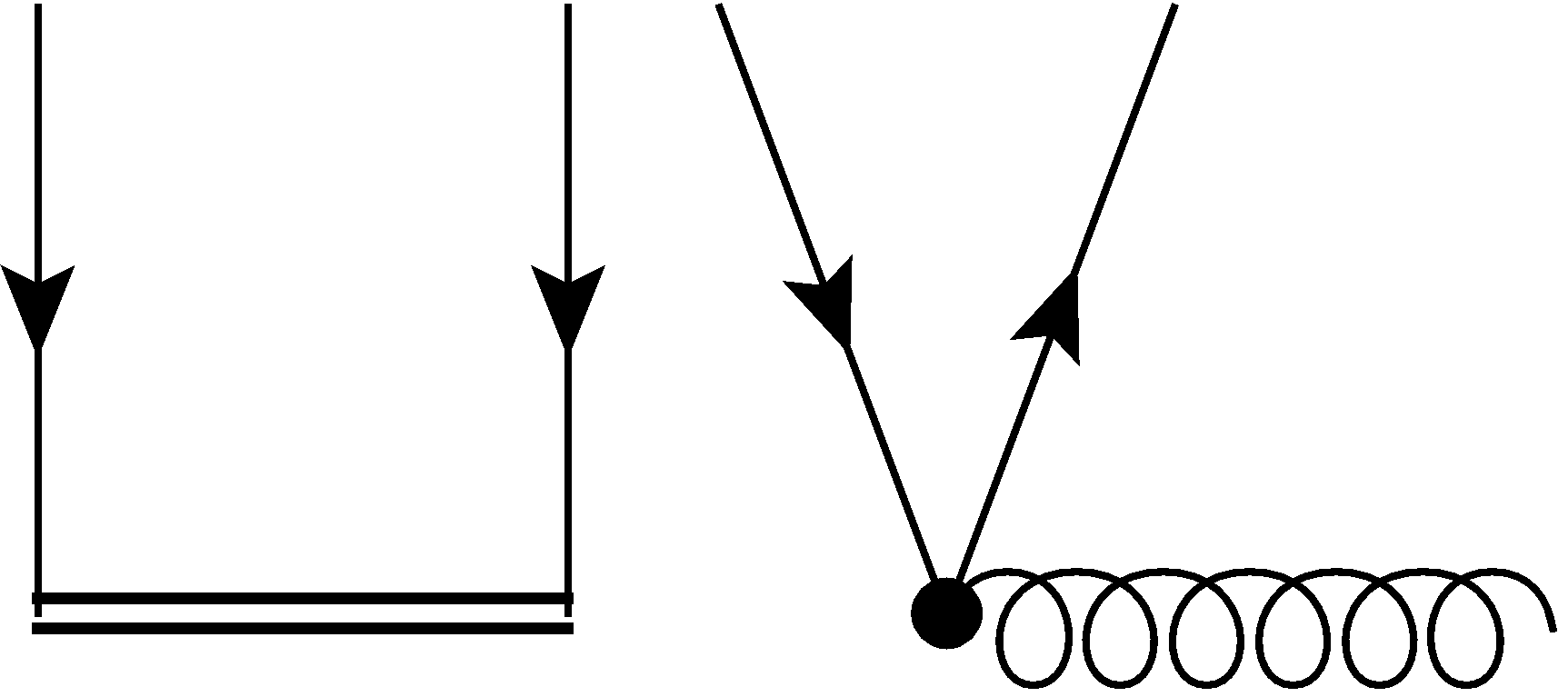} } 
        & $ P(ij,k) (r_0)_{ij} \Bar{\Theta}^{a}_{k} $
        & $ P(ij,k) \left[ \delta_{J_{ij} 0} \delta_{J_{ijk} j_k} \, (r_0)_{ij}(J=0) \, \Bar{\Theta}^{a}_{k}(K) 
        \right] $
        \\
        \vskip 4mm
        % ph-hh
        \centering\raisebox{-0.5\height}
        { \includegraphics[width=0.25\columnwidth]{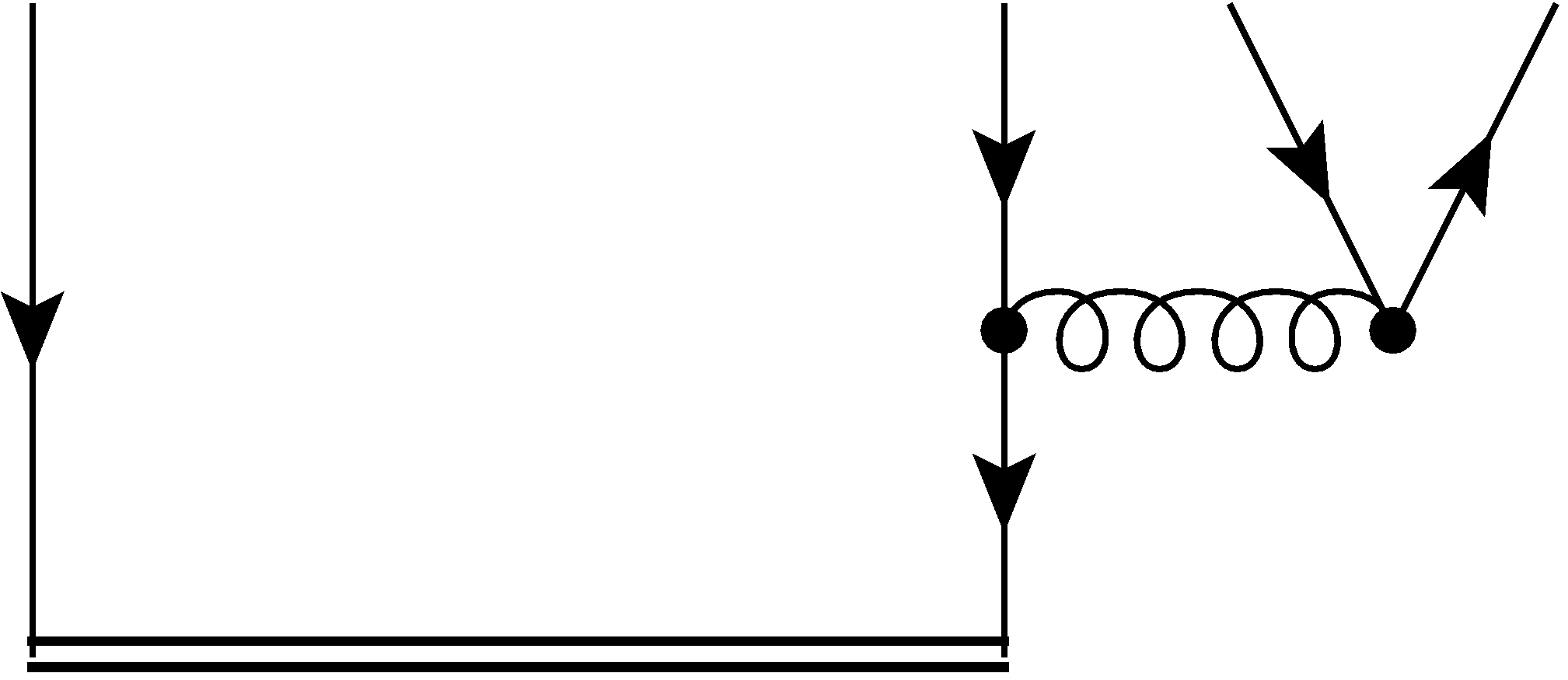} } 
        & $ - P(ij,k) \sum_m (r_0)_{mk} \Bar{\Theta}_{ij}^{am}$ 
        & $ - \frac{ \hat{J}_{ijk} }{ \hat{j}_a \hat{j}_k } P(ij,k) \, \sum_{ m J_{am} } 
        \, (-1)^{ K + J_{ij} + J_{am} }
        \, \hat{J}_{am}^{2} \delta_{j_k j_m} $ \\
        & & $ \qquad \times \,
        \begin{Bmatrix}
            J_{ij} & j_k & J_{ijk} \\ j_a & K & J_{am}
        \end{Bmatrix}
        (r_0)_{mk}(J=0) \Bar{\Theta}_{ij}^{am}(K, J_{am}, J_{ij} )$ 
        \\
        \vskip 4mm
        % pp
        \centering\raisebox{-0.5\height}
        { \includegraphics[width=0.25\columnwidth]{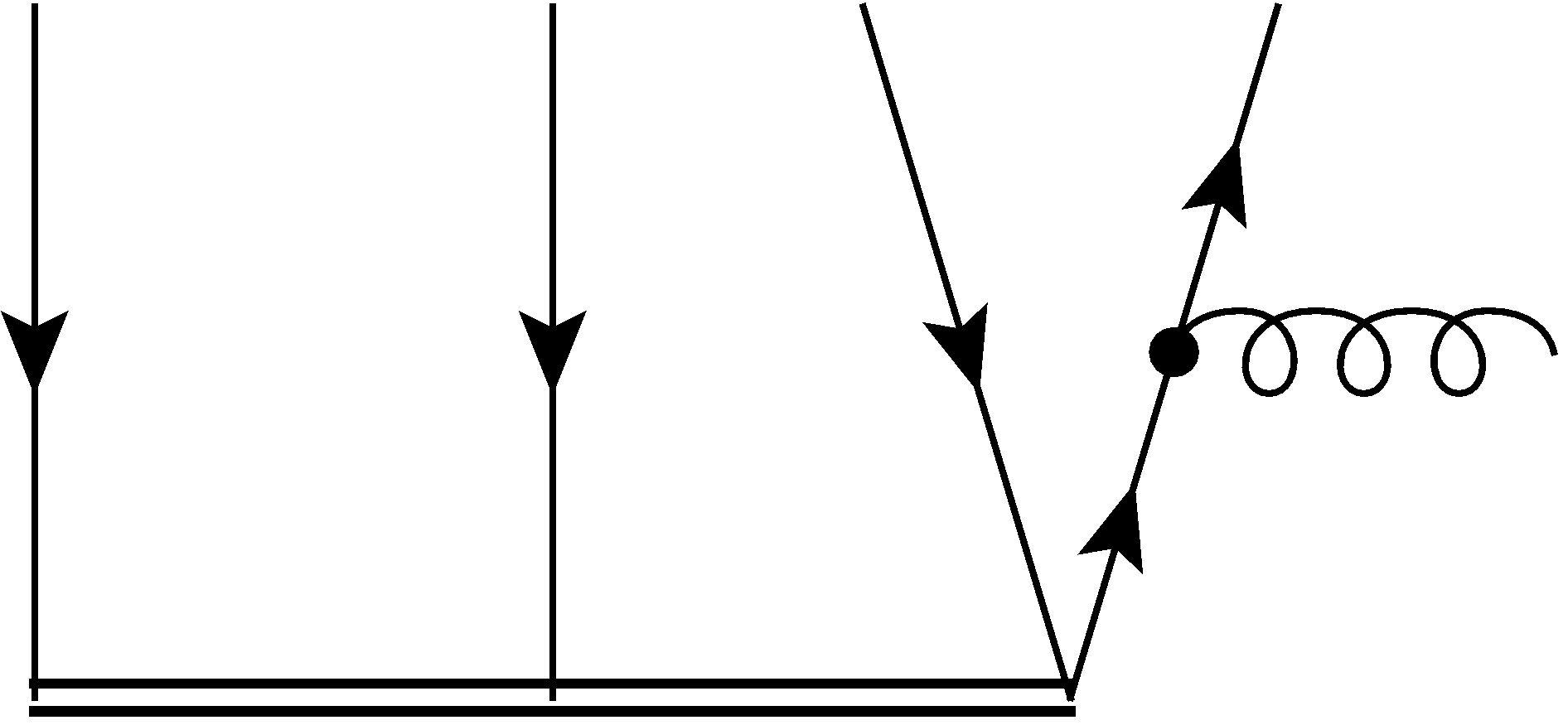} }
        & $ \sum_{e} (r_0)_{ijk}^{e}  \Bar{\Theta}^{a}_{e} $
        & $ \sum_{e} \delta_{j_e J_{ijk} }(r_0)_{ijk}^{e}( J_{ij} )  \Bar{\Theta}^{a}_{e}(K) $
        \\
        \vskip 4mm
        % hh
        \centering\raisebox{-0.5\height}
        { \includegraphics[width=0.25\columnwidth]{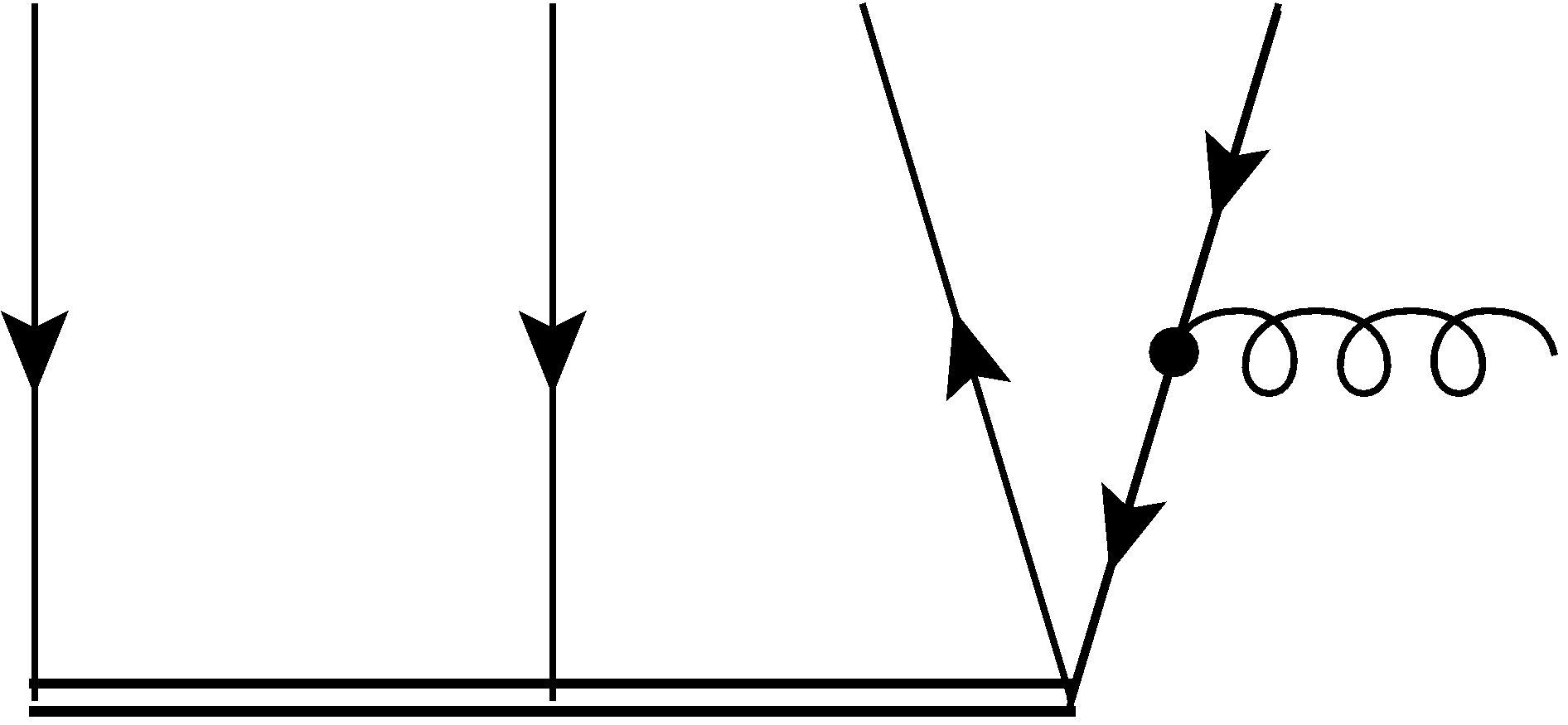} }
        & $ - P(ij,k) \sum_{m} (r_0)^{a}_{ijm} \Bar{\Theta}^{m}_{k} $
        & $ - P(ij,k) \sum_{m} \,
        (-1)^{ j_m + J_{ij} + J_{ijk} + K} \hat{J}_{ijk} \hat{j}_{m} $ \\
        & &
        $ \qquad \times \,
        \begin{Bmatrix}
            j_k & J_{ij} & J_{ijk} \\
            j_a & K & j_m
        \end{Bmatrix} \,
        (r_0)^{a}_{ijm}( J_{ij} ) \Bar{\Theta}^{m}_{k}(K) $
    \end{tabular}
    \label{tab: 2pr LIT diagrams}
\end{table*}

\begin{table*}[h]
    \caption{
    Coupled-cluster diagrams for the left pivot $\mathbf{S}^{L}$ in the 2PR-LIT approach.
    Both ordinary and reduced expressions are reported. 
    The g.s.~ amplitudes $\hat{L}_{0}$ of the 2PR nucleus are represented by double horizontal lines.
    }
    \begin{tabular}
    {>{\raggedright} m{4cm} m{4cm} m{8cm} }
        Diagram & Uncoupled expression & Coupled expression \\
        \hline
        \noalign{\vskip 2mm}   
        % hh
        \centering\raisebox{-0.5\height}
        { \includegraphics[width=0.25\columnwidth]{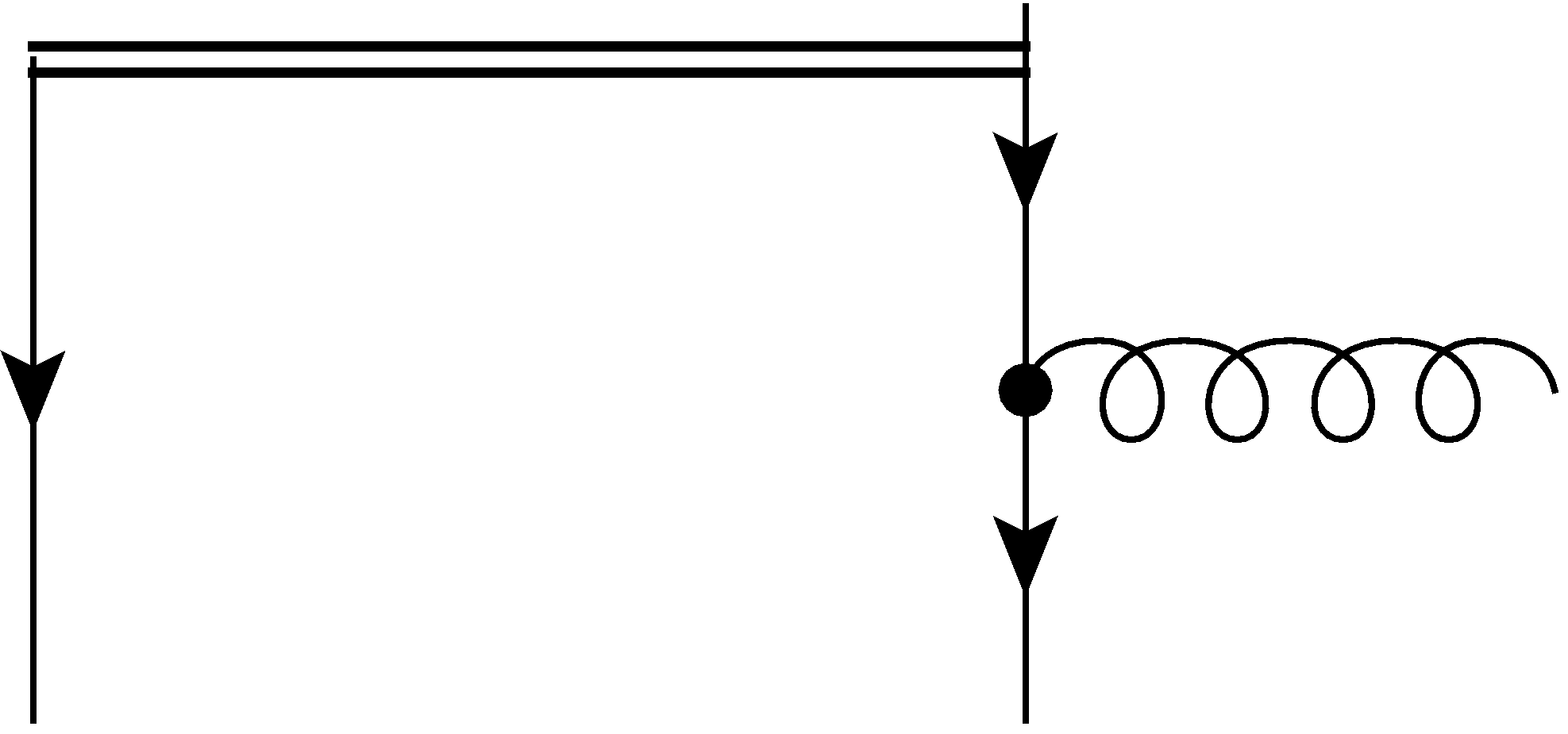} } & 
        $-P(ij) \sum_m (l_0)^{im} \overline{ \Theta^\dagger }_{m}^{j}$ & 
        $-P(ij) (-1)^{ j_i - j_j -K} \, \sum_m (l_0)^{im}(J_{im}=0) \overline{ \Theta^\dagger }_{m}^{j}(K)$
        \\
        \vskip 4mm
        % hp
        \centering\raisebox{-0.5\height}
        { \includegraphics[width=0.25\columnwidth]{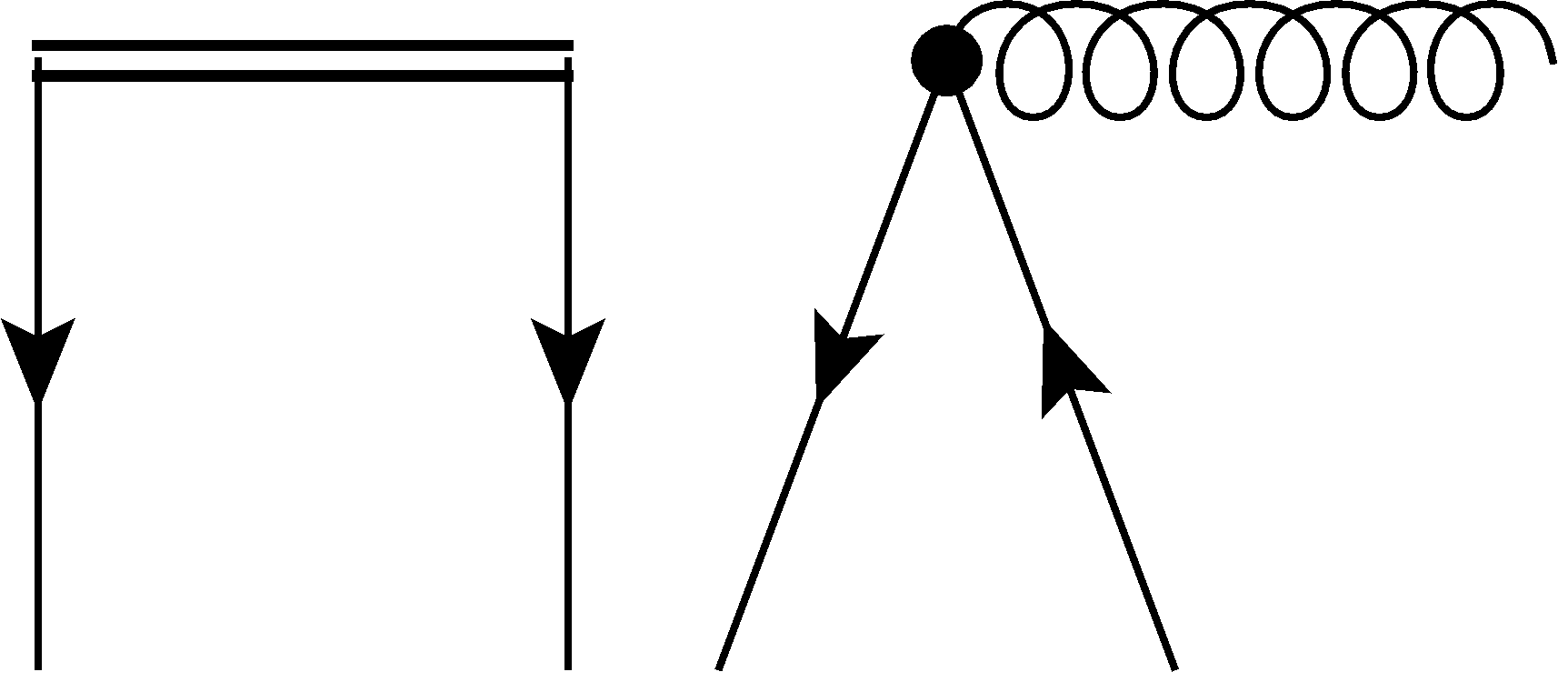} } & $ P(ij,k) (l_0)^{ij} \overline{\Theta^\dagger}_{a}^{k} $
        & $ P(ij,k) \left[ \delta_{J_{ij} 0} \delta_{J_{ijk} j_k} \, (l_0)^{ij}(J=0) \, \overline{\Theta^\dagger}_{a}^{k}(K) 
        \right] $
        \\
        \vskip 4mm
        % ph
        \centering\raisebox{-0.5\height}
        { \includegraphics[width=0.25\columnwidth]{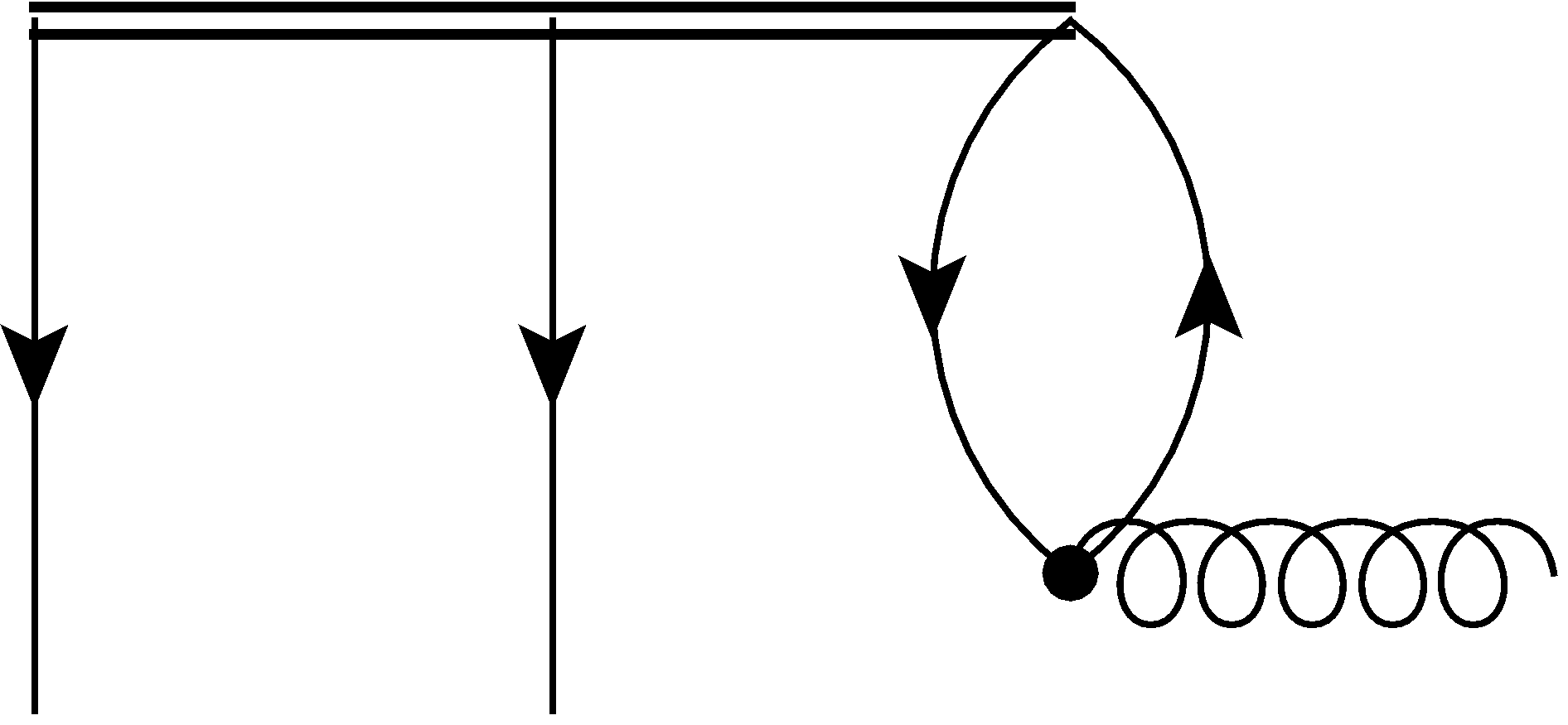} } & $ P(ij) \sum_{dm} (l_0)_{d}^{ijm} \overline{\Theta^\dagger}_{m}^{d}$
        & $ P(ij) \sum_{dm}
        \frac{ \hat{j}_d \hat{j}_m }{ \hat{K} } \,
        (-1)^{j_m + K - j_d} \,
        (l_0)_{d}^{ijm}(J_{ij}=K) \, \overline{\Theta^\dagger}_{m}^{d}(K) $
        \\
        \vskip 4mm
        % hp-hh
        \centering\raisebox{-0.5\height}
        { \includegraphics[width=0.25\columnwidth]{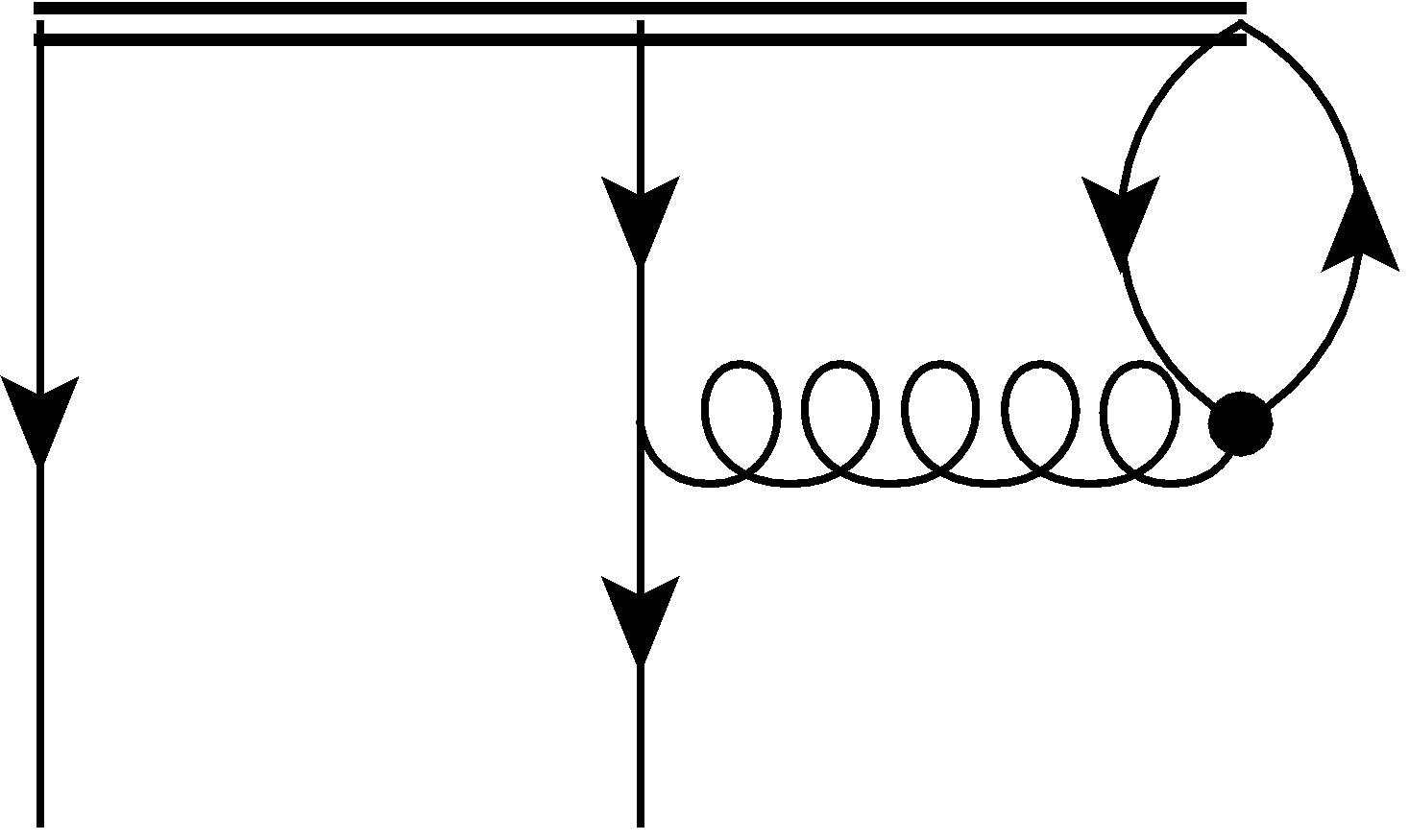} } &
        $ \frac{1}{2} P(ij) \sum_{dmn} (l_0)^{mni}_{d} \overline{\Theta^\dagger}^{jd}_{mn} $
        &
        $  \frac{1}{2}  P(ij)\sum_{dmn}
        \, (-1)^{j_i + j_d + J_{mn} + K }
        \, \hat{j}_e \hat{J}_{id} \hat{J}_{mn}
        $
        \\ \vskip 4mm
        & &
        $
        \qquad \times \,
        \begin{Bmatrix}
            J_{id} & j_d & j_i \\
            j_j & K & J_{mn}
        \end{Bmatrix} \,
        (l_0)^{mni}_{d}(J_{mn}) \overline{\Theta^\dagger}^{jd}_{mn}(K, J_{jd}, J_{mn} ) $
        \\
        \vskip 4mm
        % pp
        \centering\raisebox{-0.5\height}
        { \includegraphics[width=0.25\columnwidth]{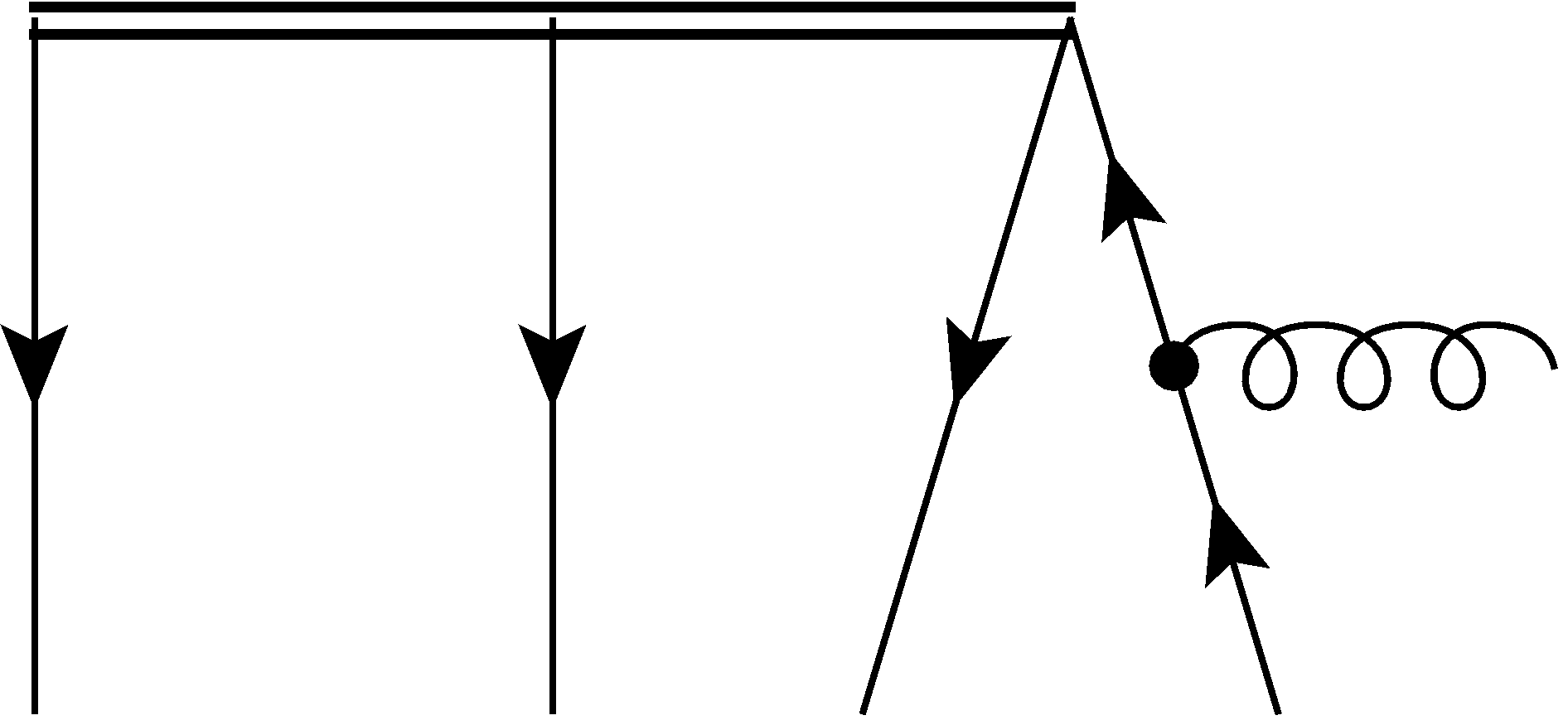} } & $ \sum_{d} (l_0)^{ijk}_{d}  \overline{\Theta^\dagger}_{a}^{d} $
        & $ \sum_{d} \delta_{j_d J_{ijk} } (l_0)^{ijk}_{d}( J_{ij} )  \overline{\Theta^\dagger}_{a}^{d}(K) $
        \\
        \vskip 4mm
        % hh
        \centering\raisebox{-0.5\height}
        { \includegraphics[width=0.25\columnwidth]{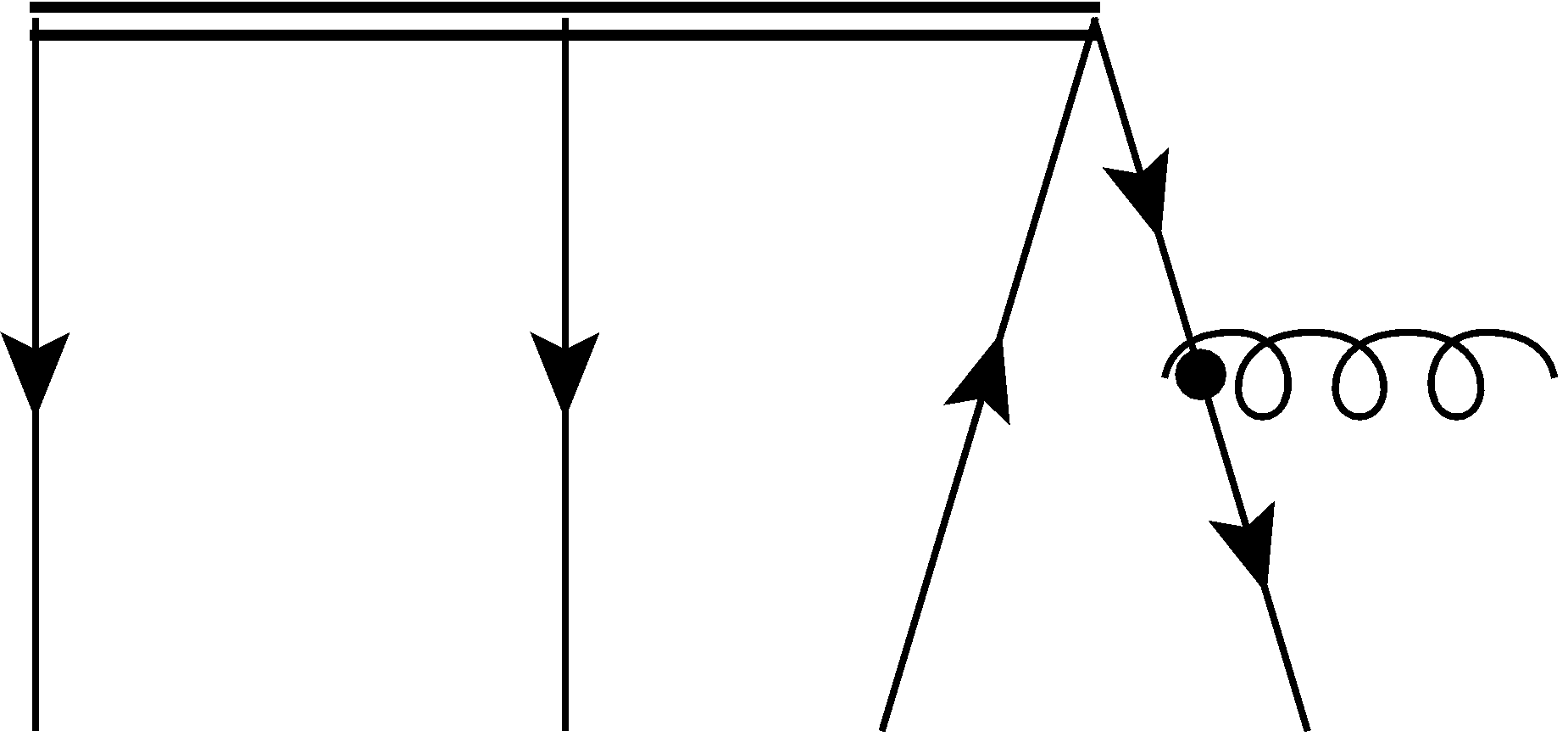} } & $ - P(ij,k) \sum_{m} (l_0)_{a}^{ijm} \overline{ \Theta^\dagger }_{m}^{k} $
        & $ - P(ij,k) \sum_{m} \,
        (-1)^{ j_m + J_{ij} + J_{ijk} + K} \hat{J}_{ijk} \hat{j}_{m} $ \\
        & &
        $ \qquad \times \,
        \begin{Bmatrix}
            j_k & J_{ij} & J_{ijk} \\
            j_a & K & j_m
        \end{Bmatrix} \,
        (l_0)_{a}^{ijm}( J_{ij} ) \overline{ \Theta^\dagger }_{m}^{k}(K) $
    \end{tabular}
    \label{tab: 2pr LIT diagrams Left}
\end{table*}

\section{Partial norms}
\label{sec: Partial norms}
We define partial norms for the EOM amplitudes following Refs.~\cite{Jansen2013,Jansen2011}.
The 2PR norms are given in terms of spherical amplitudes by
\begin{align}
    & n(0p {\text -} 2h) = \frac{1}{2}
    \sum_{ij} (r_{ij}(J))^{2}, \\
    & n(1p {\text -} 3h) = \frac{1}{6}
    \sum_{ijka} \hat{j}_{a}^{2} \, \big( r_{ijk}^{a}(J,J_{ij},J_{ijk}) \big)^{2},
\end{align}
with $n(0p {\text -} 2h) + n(1p {\text -} 3h) = 1$.
In the same way, one can introduce $1p$-$1h$ and $2p$-$2h$ norms for closed-shell EOM,
\begin{align}
    & n(1p {\text -} 1h) =
    \sum_{ai} \hat{j}_{a}^{2} \, (r_{i}^{a}(J))^{2}, \\
    & n(2p {\text -} 2h) = \frac{1}{4}
    \sum_{abij} \hat{J}_{ab}^{2} \, \big( r_{ij}^{ab}(J,J_{ab},J_{ij}) \big)^{2},
\end{align}
where $n(1p {\text -} 1h) + n(2p {\text -} 2h) = 1$.
Spherical $r_{i}^{a}$ and $r_{ij}^{ab}$ amplitudes are defined e.g. in Ref.~\cite{Hagen2014Review}.

\bibliography{bibliography.bib} 

\end{document}